\documentclass[jgrga]{agutex}

\usepackage[dvipdf]{graphicx}

\usepackage{amsmath}

 \setkeys{Gin}{draft=false}

\newcommand{\figsquash}{\ifjdraft\baselineskip=5pt\fi}

\authorrunninghead{J. D. NICHOLS}

\titlerunninghead{JOVIAN M-I COUPLING AND MAGNETODISC}

\authoraddr{J. D. Nichols,
Department of Physics and Astronomy,
University of Leicester, Leicester, LE1~7RH, UK 
(jdn@ion.le.ac.uk)}

\begin{document}

%
%

\title{Magnetosphere-ionosphere coupling in Jupiter's middle magnetosphere: computations including a self-consistent current sheet magnetic field model}

%
%

\author{J.~D.~Nichols}
\affil{Department of Physics and Astronomy, University of Leicester, UK}

%
%

\begin{abstract}
In this paper we consider the effect of a self-consistently computed magnetosdisc field structure on the magnetosphere-ionosphere coupling current system at Jupiter.  We find that the azimuthal current intensity, and thus the stretching of the magnetic field lines, is dependent on the magnetosphere-ionosphere coupling current system parameters, i.e.\ the ionospheric Pedersen conductivity and iogenic plasma mass outflow rate.  Overall, however, the equatorial magnetic field profiles obtained are similar in the inner region to those used previously, such that the currents are of the same order as previous solutions obtained using a fixed empirical equatorial field strength model, although the outer fringing field of the current disc acts to reverse the field-aligned current in the outer region.  We also find that, while the azimuthal current in the inner region is dominated by hot plasma pressure, as is generally held to be the case at Jupiter, the use of a realistic plasma angular velocity profile actually results in the centrifugal current becoming dominant in the outer magnetosphere.  In addition, despite the dependence of the intensity of the azimuthal current on the magnetosphere-ionosphere coupling current system parameters, the location of the peak field-aligned current in the equatorial plane also varies, such that the ionospheric location remains roughly constant.  It is thus found that significant changes to the mass density of the iogenic plasma disc are required to explain the variation in the main oval location observed using the Hubble Space Telescope. 
\end{abstract}
%
%

\begin{article}

%

\section{Introduction}
\label{sec:intro}

\begin{figure}[t]
 \noindent\includegraphics[width=19pc]{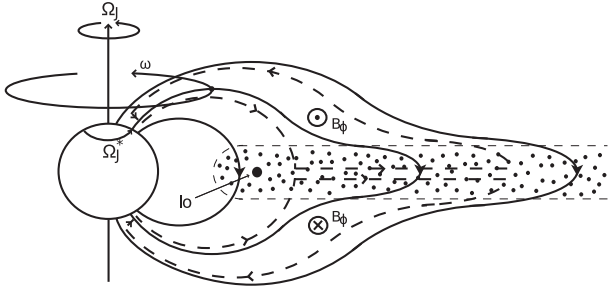}
\caption{
Sketch of a meridian cross section through Jupiter's inner and middle magnetosphere, showing the principal physical features involved. The arrowed solid lines indicate magnetic field lines, the arrowed dashed lines the magnetosphere-ionosphere coupling current system, and the dotted region the rotating disc of outflowing plasma.  After \cite{cowley01}.
}
\label{fig:miccs}
\end{figure}

The dynamics of Jupiter's middle magnetosphere are dominated by planetary rotation coupled with the centrifugally-driven outflow of plasma from the volcanic moon Io, which orbits at \ensuremath{\sim}5.9~\ensuremath{\mathrm{R_J}} (where \ensuremath{\mathrm{R_J}} is the equatorial radius of Jupiter equal to 71,373~km) and liberates \ensuremath{\sim}1000~\ensuremath{\mathrm{kg\;s^{-1}}} of sulphur and oxygen into a torus surrounding the satellite's orbit \citep[e.g.][]{siscoe81, vasyliunas83, khurana93, delamere03a, dols08a}.  Subrotation of outflowing equatorial plasma leads to the bend-back of magnetic field lines out of the meridian planes and the formation of a large-scale magnetosphere-ionosphere (M-I) coupling current system, illustrated schematically in Figure~\ref{fig:miccs} \citep[e.g.][]{hill79, hill01, pontius97, cowley01, nichols03, nichols04, nichols05}.  The current system, which consists of an equatorward-directed Pedersen current in the ionosphere and a radial current in the equatorial plane joined in the inner region by an upward-directed field-aligned (Birkeland) current and closed in the outer region by a downward-directed current, communicates drag from the atmospheric neutrals to the equatorial plasma.  The upward field-aligned component of this current system, associated with downward-precipitating electrons, is thought to be the cause of Jupiter's main auroral oval, which is the most significant of Jupiter's various ultraviolet (UV) auroral forms \citep{grodent03b, clarke04, nichols09b}.  \\

This jovian M-I coupling current system was studied originally by \cite{hill79}, who calculated the plasma angular velocity profile employing a dipole planetary magnetic field, and the theory was later generalised to include a realistic current sheet magnetic field (magnetodisc) model by \cite{pontius97}. The link with the main oval auroral emission was realised later by \cite{hill01}, who again used a theoretical angular velocity profile calculated using a dipole field, and \cite{cowley01}, who used empirical plasma angular velocity and current sheet magnetic field profiles.  \cite{cowley02} computed using Hill-Pontius theory the plasma angular velocity and current profiles using both dipole and current sheet field models, and showed that the stretching of the equatorial middle magnetosphere field lines associated with the current sheet dramatically alters the magnitude and location of the auroral field-aligned currents.  The effects of two poorly-constrained system parameters, the effective ionospheric Pedersen conductance \ensuremath{\Sigma_P^*}, and the plasma mass outflow rate \ensuremath{\dot{M}} were then studied in detail by \cite{nichols03}, and \cite{nichols04} went on to examine the effect of self-consistent modulation of the ionospheric Pedersen conductance due to auroral electron precipitation.  \cite{nichols05} and \cite{ray10a} have since studied the effect of field-aligned voltages, and \cite{cowley07} examined the effects of solar wind-induced expansions and contractions of the planet's magnetosphere.  In addition, the modulation of the current system by diurnal variation of the ionospheric Pedersen conductance caused by solar illumination has been studied by \cite{tao10a}.  Most recently, \cite{nichols11a} applied approximations derived by \cite{nichols03} for the jovian M-I coupling current system to the cases of rapidly rotating, strongly illuminated Jupiter-like exoplanets, the radio emissions from which may offer a novel method for detecting such objects.\\

\begin{figure}
 \noindent\includegraphics[width=19pc]{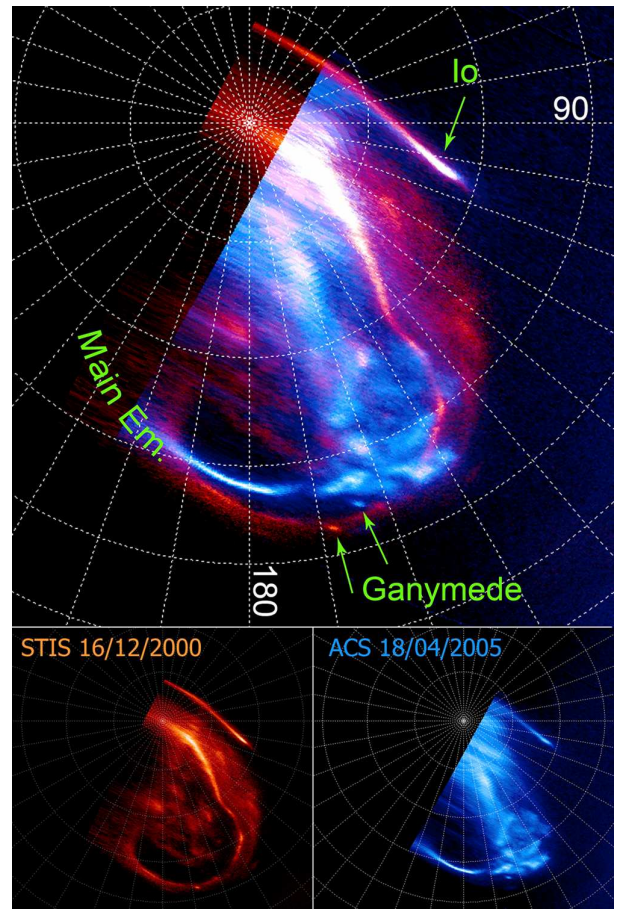}
\caption{
Top panel: Superposition of the polar projection of two images of Jupiter's northern aurora obtained with HST more than four years apart.  The  red image was obtained with the Space Telescope Imaging Spectrograph (STIS) in December 2000, and the blue image was obtained with the Advanced Camera for Surveys (ACS) camera in April 2005. The 90\ensuremath{^\circ{}} and 180\ensuremath{^\circ{}} System III meridians have been highlighted on a 10° spaced grid. Green arrows point to the footprints of Ganymede and Io, and the main emission has also been marked in green.  Bottom panels: Individual polar projections using the same longitude system as in top panel. From \cite{grodent08}. 
}
\label{fig:g08} 
\end{figure}

A key limitation of the preceding studies, however, is that they have all employed fixed magnetic field models as the basis for the computations, be they dipole or current sheet in form.  However, as shown in Figure~\ref{fig:g08} (reproduced from \cite{grodent08}), the main oval has been observed to shift in latitude by up to \ensuremath{\sim}3$^\circ$ when comparing images obtained a number of years apart.  This shift in the latitude of the main oval was accompanied by a similar shift of \ensuremath{\sim}2$^\circ$ in the latitude of the footprint of Ganymede, such that \cite{grodent08} attributed the shift in the main oval to a change in the intensity of the azimuthal current, which affects the mapping between ionosphere and equator, rather than simply a shift across L-shells of the field-aligned current, as is predicted would occur for different values of \ensuremath{\Sigma_P^*} and \ensuremath{\dot{M}} \citep[e.g.][]{nichols03}. \cite{caudal86} showed that the stretching of Jupiter's middle magnetosphere field lines is caused in part by the centrifugal force of iogenic plasma, a quantity specifically associated with the iogenic plasma mass density and angular velocity profile.  He constructed a model for Jupiter's magnetodisc by modifying a terrestrial storm-time ring current model \citep{lackner70a}, and employing inputs based on Voyager plasma temperature and density observations, along with a fixed angular velocity profile given by  the theory of \cite{hill79}, computed using a dipole field model.  \cite{caudal86} did note the inconsistency in employing an angular velocity profile calculated using a dipole field, but while \cite{pontius97} and \cite{cowley02} showed that the equatorial plasma angular velocity profile is relatively insensitive to the field model used, the latter authors showed that the resulting auroral currents are very sensitive to the model employed. In addition, in calculating the iogenic plasma angular velocity profile \cite{caudal86} used a corotation breakdown scale distance $\rho_H$ (termed the `Hill distance') of 20~\ensuremath{\mathrm{R_J}} in conformity with the value deduced by \cite{hill80}, and in his model the plasma angular velocity thus falls to \ensuremath{\sim}17\% of rigid corotation by 60~\ensuremath{\mathrm{R_J}}.  However, observational studies such as \cite{kane95} have reported that the plasma angular velocities remain at \ensuremath{\sim}50\% of rigid corotation out to \ensuremath{\sim}60~\ensuremath{\mathrm{R_J}}, and \cite{hill01} later revised his estimate of $\rho_H$ to 30~\ensuremath{\mathrm{R_J}}, such that the middle magnetosphere plasma angular velocities employed by \cite{caudal86}, and thus the centrifugal force imparted by the iogenic plasma, are somewhat lower than realistically expected.  The purpose of the present paper is thus as follows.  First, we incorporate the calculation of the plasma angular velocity profile using Hill-Pontius theory into the model of \cite{caudal86}, such that the resulting magnetosphere-ionosphere currents are computed using values of the equatorial magnetic field self-consistent with the plasma angular velocity profile.  Second, in doing so we will update the model results of \cite{caudal86} using more realistic plasma parameters, including values obtained from Galileo data. We then examine the effect of the ionospheric Pedersen conductance and iogenic plasma mass outflow rate in order to compare results with the previous modeling work of \cite{nichols03}, and examine whether variations of these parameters may be responsible for the changing auroral locations observed by \cite{grodent08}.

\section{Theoretical background}
\label{sec:theory}

\subsection{M-I coupling current system equations}
\label{sec:miccs}

We begin by discussing the equations governing the jovian M-I coupling current system.  The system has been discussed in depth previously \citep[e.g.][]{hill79, pontius97,cowley02, nichols03,nichols04,nichols05}, such that only a brief outline is given here.  We first assume axisymmetry, such that Jupiter's poloidal field can be described with the use of a flux function $F(\rho,z)$, related to the Euler potential $\alpha$ used in Section~\ref{sec:disc}, and from which the magnetic field can be computed via

\begin{equation}
	\mathbf{B} = \left(\frac{1}{\rho}\right)\nabla F \times \hat{\varphi}\;\;,
	\label{eq:bf}
\end{equation}

\noindent where $\rho$ is the perpendicular distance from the magnetic axis, $z$ is the distance along this axis from the magnetic equator, and $\varphi$ is the azimuthal angle.  Mapping between the equator and ionosphere is then easily achieved by writing $F_e=F_i$, where subscript `e' refers to the equator and `i' refers to the ionosphere.  Although the magnetic field model discussed in Section~\ref{sec:disc} does compute the field at all latitudes and altitudes down to 1~\ensuremath{\mathrm{R_J}}, such that in principle it provides ionospheric field values, \cite{caudal86} pointed out that the simple treatment of the magnetopause currents renders the model invalid at latitudes above \ensuremath{\sim 50^\circ}, and in addition the cartesian grid used here is too coarse to be useful in this region.  However, the ionospheric field is overwhelmingly dominated by the planetary dipole, such that in common with previous works we assume the field in the ionosphere is purely dipolar.  The ionospheric flux function is then given by 

\begin{equation}
	F_i=B_J\rho_i^2=B_JR_J^2\sin^2\theta_i\;\;,
	\label{eq:fi}
\end{equation}

\noindent where $B_J$ is the dipole equatorial magnetic field strength (equal to 426, 400 nT in conformity with the VIP 4 internal field model of \cite{connerney98}), $R_J$ is Jupiter's radius (equal to 71 373 km), $\rho_i$ is the perpendicular distance from the magnetic axis, and $\theta_i$ represents magnetic co-latitude.  \\

The application of Newton's second law to an axisymmetric radially-outward steady flow of plasma from the torus source yields the `Hill-Pontius' differential equation for the plasma angular velocity $\omega$, given by

\begin{equation}
	\frac{\rho_e}{2}\frac{d}{d\rho_e}\left(\frac{\omega}{\Omega_J}\right)+\left(\frac{\omega}{\Omega_J}\right)=\frac{4\pi \Sigma_P^*F_e|B_{ze}|}{\dot{M}}\left(1-\frac{\omega}{\Omega_J}\right)\;\;,
	\label{eq:hp}
\end{equation}

\noindent where $\rho_e$ represents equatorial radial distance, \ensuremath{\Omega_J} is the planet's angular velocity equal to $1.76\times10^{-4}\mathrm{\;rad\;s^{-1}}$, and $|B_{ze}|$ is the magnitude of the north-south magnetic field threading the equatorial plane.  Note that the effective Pedersen conductance \ensuremath{\Sigma_P^*} (here defined for one hemisphere) is reduced from the true value \ensuremath{\Sigma_P} by $\ensuremath{\Sigma_P^*}=(1-k)\ensuremath{\Sigma_P}$, where the parameter $k$ represents the reduction of the angular velocity of the neutral atmosphere ($\Omega_J^*$) from rigid corotation ($\Omega_J$) via `slippage' \citep{huang89,millward05}, such that  $(\Omega_J-\Omega_J^*)=k(\Omega_J-\omega)$.  The value of $k$ is somewhat uncertain, so in common with previous studies we take $k=0.5$, although we note that in reality this  approach may be an oversimplification \citep{smith09a,tao09a}.  The quantities $F_e$ and $|B_{ze}|$ are obtained from the magnetodisc model discussed in Section~\ref{sec:disc}, such that Eq.~\ref{eq:hp} is solved numerically to obtain the equatorial plasma angular velocity profile. \\  

We now discuss the equations which describe the resulting magnetosphere-ionosphere coupling currents.  First, the equatorward-directed height-integrated Pedersen current $i_P$ is given by

\begin{equation}
	i_P=2 \ensuremath{\Sigma_P^*}B_J\Omega_J\rho_i\left(1-\frac{\omega}{\Omega_J}\right)\;\;,
	\label{eq:ip}
\end{equation}

\noindent where we have taken the ionospheric field to be vertical and equal to $2B_J$ in strength (an approximation valid to within \ensuremath{\sim}5\% in our region of interest \citep{nichols03}).  Current continuity and the assumption of north-south symmetry then yields for the equatorial radial current integrated across the width of the current sheet $i_\rho$

\begin{equation}
	\rho_e i_\rho=2\rho_i i_P\;\;.
	\label{eq:irhoip}
\end{equation}

\noindent Recalling that $F_i=F_e$, we have from Eqs.~\ref{eq:ip}, ~\ref{eq:irhoip} and \ref{eq:fi} 

\begin{equation}
	i_\rho=\frac{4 \ensuremath{\Sigma_P^*}F_e\Omega_J}{\rho_e}\left(1-\frac{\omega}{\Omega_J}\right)\;\;,
	\label{eq:irho}
\end{equation}

\noindent such that the total radial current integrated in azimuth $I_\rho$ is

\begin{equation}
	I_\rho=2\pi\rho_ei_\rho=8\pi \ensuremath{\Sigma_P^*}\Omega_J F_e \left(1-\frac{\omega}{\Omega_J}\right)\;\;,
	\label{eq:totip}
\end{equation}

\noindent which is equal to twice the azimuth-integrated equatorward-directed Pedersen current $I_P$ flowing in each hemisphere.  The field-aligned current density at the top of the ionosphere \ensuremath{j_{\|i}} is then computed from the divergence of either total field-perpendicular current, such that, in terms of the radial current

\begin{equation}
	\ensuremath{j_{\|i}}=\frac{B_J}{2\pi\rho_e|B_{ze}|}\frac{dI_\rho}{d\rho_e}\;\;.
	\label{eq:jpari}
\end{equation}

\subsection{Magnetodisc field model}
\label{sec:disc}

Whereas previous studies have specified the equatorial magnetic field profile as a fixed input to the equations discussed in Section~\ref{sec:miccs}, here we self-consistently employ the jovian magnetodisc model of \cite{caudal86}, which we note has also recently been adapted to the saturnian magnetodisc \citep{achilleos10a}.  Briefly, \emph{Caudal's}~[1986] model represents the spin-aligned magnetic field as the gradient of the Euler potentials, given generally by $\alpha$ and $\beta$, but which in the axisymmetric case are reduced to one function $\alpha(r,\theta)$, where $(r,\theta)$ are polar coordinates.  The Euler potential $\alpha$ is related to the flux function $F$ via

\begin{equation}
	F=R_J\alpha\;\;,
	\label{eq:alphaf}
\end{equation}

\noindent  such that the magnetic field components are then given e.g.\ in cylindrical coordinates by

\begin{subequations}
	\label{eq:bcomps}
	\begin{equation}
	B_\rho=-\frac{R_J}{\rho}\frac{\partial\alpha}{\partial z}\;\;,
	\label{eq:brho}
	\end{equation}
	\begin{equation}
	B_z=\frac{R_J}{\rho}\frac{\partial\alpha}{\partial\rho}\;\;,
	\label{eq:bz}
	\end{equation}
	\begin{equation}
	B_\varphi=0\;\;.
	\label{eq:bphi}
	\end{equation}

\end{subequations}

\noindent Note that this assumes that the poloidal magnetodisc structure is unaffected by the small azimuthal field generated by the equatorial radial component of the M-I coupling current system.  \cite{caudal86} considered the momentum equation for a rotating plasma, i.e.\ 

\begin{equation}
	\mathbf{j}\times\mathbf{B}=\nabla P - d \omega^2 \rho \underline{\hat{\rho}}\;\;,
	\label{eq:mom}
\end{equation}

\noindent where $\mathbf{j}$ is the current density, $P$ is the plasma pressure and $d$ is the plasma mass density.  From this he derived the differential equation

\begin{equation}
	\frac{\partial^2\alpha}{\partial r^2}+\frac{(1-x^2)}{r^2}\frac{\partial^2\alpha}{\partial x^2}
	=-g(r,x,\alpha)\;\;,
	\label{eq:alphag}
\end{equation}

\noindent where $x = \cos\theta$ and the function $g$, which is derived from the plasma pressure and angular velocity distributions, is related to the azimuthal current density $j_\varphi$ via

\begin{equation}
	g = \mu_0j_\varphi\rho \;\;.
	\label{eq:jg}
\end{equation}

\noindent Function $g$ comprises two summed components, representing contributions from the hot \ensuremath{\sim}30~keV \citep{krimigis81} and cold \ensuremath{\sim}100~eV \citep{mcnutt81a,frank02a} plasma populations in Jupiter's magnetosphere.  For the hot plasma population (subscript `h') the pressure gradient dominates the centrifugal force, such that the latter is neglected and $g$ is given by 

\begin{equation}
	g_h=\mu_\circ\left(\frac{r}{R_J}\right)^2(1-x^2)\frac{dP_h}{d\alpha}\;\;,
	\label{eq:ghot}
\end{equation}

\noindent while the source function for the cold population (subscript `c') includes the centrifugal force, such that

\begin{equation}
	g_c=\mu_\circ\left(\frac{\rho}{R_J}\right)^2\exp\left(\frac{\rho^2-\rho_\circ^2}{2l^2}\right)\left[\frac{dP_{c\,\circ}}{d\alpha}+\frac{P_{c\,\circ} R_J}{l^2 |B_{ze\circ}|}\right] \;\;,
	\label{eq:gcold}
\end{equation}

\noindent where $\rho_\circ{}$, $P_{c\,\circ}$ and $B_{ze\circ{}}$ are the values at the equatorial crossing point of the field line. Quantity $l$ in Eq.~\ref{eq:gcold} represents the centrifugal equatorial confinement scale height of the cold plasma, given for a singly-ionised, monoionic population with temperature $T_c$ and ion mass $m$ by 

\begin{equation}
	l=\left(\frac{2kT_c}{\omega^2 m}\right)^{\frac{1}{2}}\;\;,
	\label{eq:l}
\end{equation}

\noindent where $k$ is Boltzmann's constant equal to $1.38\times 10^{-23}\mathrm{\;J\;K^{-1}}$.  The plasma pressure $P$ is given in general by the ideal gas law for a singly-ionised plasma

\begin{equation}
	P = \frac{2NkT}{V}\;\;,
	\label{eq:p}
\end{equation}

\noindent where $N$ is the number of ions per Wb and $V(\alpha)$ is the volume of the unit flux tube, given for the hot plasma by 

\begin{equation}
	V_h=\int\frac{ds}{B}\;\;,
	\label{eq:vh}
\end{equation}

\noindent and for the centrifugally confined cold plasma by

\begin{equation}
	V_c=\int\exp{\left(\frac{\rho^2-\rho_\circ^2}{2l^2}\right)}\frac{ds}{B}\;\;,
	\label{eq:vc}
\end{equation}

\noindent where the exponential term in equation~\ref{eq:vc} represents the centrifugal confinement pressure of the cold plasma.  All the physical properties of the plasma are thus represented in $P$ and $l$, and \cite{caudal86} used Voyager observations \citep{bagenal81a,connerney81,krimigis81,mcnutt81a,siscoe81} to provide suitable values.  Specifically, he took for the hot plasma 

\begin{equation}
	P_h(\alpha) 
	  \begin{cases}
		   = \displaystyle{\frac{3.0\times 10^7}{V_h(\alpha)}} & \text{if } \rho_\circ{} \geq 7.5\;\mathrm{R_J} \\
		   \propto \rho_\circ{} & \text{if } \rho_\circ{} < 7.5\;\mathrm{R_J} \;\;,
	  \end{cases}
	\label{eq:ph}
\end{equation}

\noindent a form which we also employ here.  For the cold plasma he employed equations~\ref{eq:l}, \ref{eq:p}, and \ref{eq:vc}, with profiles for $N_c(\rho_\circ{})$ and $kT_c(\rho_\circ{})$ derived from Voyager data, i.e.\

\begin{equation}
	N_c(\rho_\circ{}) = 
	  \begin{cases}
		  0 & \text{if } \rho_\circ{} < 5\;\mathrm{R_J} \\
		  10.7\times 10^{22} & \text{if } 5.7\;\mathrm{R_J} \leq \rho_\circ{} < 7\;\mathrm{R_J} \\
		  2.9\times 10^{22} & \text{if } \rho_\circ{} \geq 8\;\mathrm{R_J} \;\;,
	  \end{cases}
	\label{eq:nc}
\end{equation}

\noindent with continuity achieved through linear interpolation between these domains, and

\begin{equation}
	kT_c(\rho_\circ{}) = 
	  \begin{cases}
		  1\text{ eV} & \text{if } \rho_\circ{} = 5\;\mathrm{R_J} \\
		  35\text{ eV} & \text{if } 6\;\mathrm{R_J} \leq \rho_\circ{} < 7\;\mathrm{R_J} \\
		  10(\rho_\circ{}/R_J)\text{ eV} & \text{if } \rho_\circ{} \geq 9\;\mathrm{R_J} \;\;,
	  \end{cases}
	\label{eq:ktc}
\end{equation}

\noindent with continuity achieved here through linear interpolation of $\log(kT_c)$. For the plasma angular velocity, \cite{caudal86} employed \emph{Hill's}~[1979] solution to Eq.~\ref{eq:hp} for the dipole field, given by

\begin{align}
	&\left(\frac{\omega}{\Omega_J}\right)=
	\frac{1}{\rho^2}\exp\left[-{\rho_H}^4\left(1-\frac{1}{\rho^4}\right)\right]+\nonumber \\
	&\sqrt{\pi}\left(\frac{\rho_H}{\rho}\right)^2\exp\left[\left(\frac{\rho_H}{\rho}\right)^4\right]
	\left\{\mathrm{erf}\left[\left(\frac{\rho_H}{\rho}\right)^2\right]-\mathrm{erf}\left({\rho_H}^2\right)\right\}
	\label{eq:hillsol}
\end{align}

\noindent where $\mathrm{erf}(z)=(2/\sqrt{\pi})\int_0^ze^{-t^2}\,dt$ is the error function and the Hill distance $\rho_H$ is given by

\begin{equation}
	\left(\frac{\rho_H}{R_J}\right)=\left(\frac{2\pi\Sigma_P^*{B_J}^2{R_J}^2}{\dot{M}}\right)^\frac{1}{4}\;\;,
	\label{eq:rhoh}
\end{equation}

\noindent which, as discussed in Section~\ref{sec:intro}, \cite{caudal86} took to be equal to 20, corresponding to a value for the quotient $(\ensuremath{\Sigma_P^*}/\ensuremath{\dot{M}}) = 2.75\times 10^{-5}\;\mathrm{mho\;s\;kg^{-1}}$ (where 1~mho~=~1~siemen).  It is also worth noting here that \emph{Hill's}~[2001] revised estimate of $(\rho_H/R_J)=30$ corresponds to $(\ensuremath{\Sigma_P^*}/\ensuremath{\dot{M}}) = 1.4\times 10^{-4}\;\mathrm{mho\;s\;kg^{-1}}$.\\

The iterative analytic solution to equation~\ref{eq:alphag}, stated by \cite{caudal86} and derived explicitly by \cite{achilleos10a}, is

\begin{align}
	&\alpha_n=\alpha_\circ+(1-x^2)\sum_{n=0}^\infty\frac{P_n^{(1,1)}(x)}{2n+3} \times \nonumber \\
	&\left[r^{-n-1}\left(\int_{r_c}^ru^{n+2}g_n(u)du\right)+r^{n+2}\int_r^\infty g_n(u)u^{-n-1}du\right]\;\;,
	\label{eq:alphan}
\end{align}

\noindent where $P_n^{(a,b)}(z)$ are the Jacobi polynomials \citep[see e.g.][]{abramowitz65a}, and 

\begin{equation}
	g_n(u)=\frac{1}{h_n}\int_{-1}^1 g(r,x)P_n^{(1,1)}(x)dx\;\;,
	\label{eq:gnu}
\end{equation}

\noindent where

\begin{equation}
	h_n=\int_{-1}^1 (1-x^2)(P_n^{(1,1)}(x))^2dx\;\;.
	\label{eq:hnu}
\end{equation}

\noindent The solution is initiated with the Euler potential for a dipole field $\alpha_\circ{}$ given by

\begin{equation}
	\alpha_\circ{} = B_J R_J^2 \left(\frac{1-x^2}{r}\right)\;\;,
	\label{eq:alpha0}
\end{equation}

\noindent and proceeds by iteration according to the scheme illustrated by Figure~4 in \cite{caudal86}.  As he noted, in order to achieve convergence, after a few iterations the new values of $\alpha$ are obtained using a weighted average of $\alpha_n$ and $\alpha_{n-1}$.  In addition, at each iteration an Euler potential $\alpha_{s}$ representing the field induced by the equatorial magnetopause current as seen inside the magnetosphere is added to the solution given by equation~\ref{eq:alphan}.  It is modeled as an irrotational field of strength $B_{s}$, such that

\begin{equation}
	\alpha_{s} = -\frac{B_{s}r^2}{2R_J}(1-x^2)\;\;,
	\label{eq:alphas}
\end{equation}

\noindent and $B_{s}$ is given by

\begin{equation}
	B_{s} = 0.6 \frac{2R_J\alpha_{mp}}{R_{mp}^2}\;\;,
	\label{eq:bs}
\end{equation}

\noindent where $R_{mp}$ is the distance to the equatorial magnetopause, taken in this study to be the representative value of 85~\ensuremath{\mathrm{R_J}}, and $\alpha_{mp}$ is the value of $\alpha$ at the equatorial magnetopause.

\subsection{Application of the magnetodisc model to M-I coupling}
\label{sec:application}

In applying \emph{Caudal's}~[1986] magnetodisc model to the jovian M-I coupling current system, the major development of the model is the treatment of the cold plasma angular velocity, which we describe below.  We first discuss, however, a secondary modification of the model, concerning the input cold plasma parameter values.  As discussed in Section~\ref{sec:disc}, \cite{caudal86} employed cold plasma ion number density values (number per Weber) based on estimates over the radial range 5-9~\ensuremath{\mathrm{R_J}} calculated by \cite{bagenal81a}, who used Voyager plasma data and assumed a dipole field and an exponential distribution of plasma along the field lines.  Between 8~\ensuremath{\mathrm{R_J}} and the magnetopause \cite{caudal86} used a constant value based on the outer values of the \cite{bagenal81a} results.  More recently, \cite{frank02a} have reported thermal plasma density observations obtained by the Galileo spacecraft over a much greater radial distance, out to 100~\ensuremath{\mathrm{R_J}}, and provided the following power laws for the thermal plasma number densities

\begin{equation}
	n_c(\rho_\circ{}) = 
	  \begin{cases}
		  3.2 \times{} 10^8\; \rho_\circ{} ^{-6.90}\; \mathrm{ cm^{-3}} & \text{if } \rho_\circ{} < 20\;\mathrm{R_J} \\
		  9.8 \;\rho_\circ{} ^{-1.28}\; \mathrm{ cm^{-3}} & \text{if }  \rho_\circ{} > 50\;\mathrm{R_J}  \;\;.
	  \end{cases}
	\label{eq:fnc}
\end{equation}

\begin{figure}[t]
 \noindent\includegraphics[width=19pc]{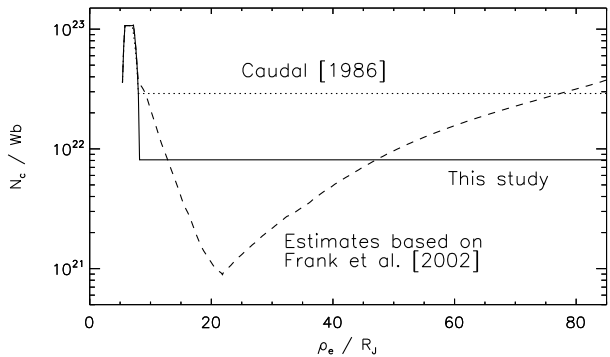}
\caption{
Plot showing the number of cold ions per Weber versus radial distance.  The solid line shows the values employed in this study, the dotted line shows the values used by \cite{caudal86}, and the dashed line shows the values estimated using the results of \cite{frank02a} along with the flux tube volume calculated using \emph{Caudal's}~[1986] model.  
}
\label{fig:nc}
\end{figure}

\noindent  Estimates of the number of ions per Weber can then be obtained by multiplying equation~\ref{eq:fnc} by the weighted flux tube volume (e.g.\ as given by equation~\ref{eq:vc}), for which the values as calculated using \emph{Caudal's}~[1986] original model may be used as reasonable estimates. The number of ions per Weber thus calculated are shown in  Figure~\ref{fig:nc}.  The dotted line (which is essentially overlaid by the solid line within 8~\ensuremath{\mathrm{R_J}}) shows the values employed by \cite{caudal86}, which we recall are unconstrained by data beyond 9~\ensuremath{\mathrm{R_J}}, while the dashed line shows the estimates using the two power laws given by \cite{frank02a} as discussed above, where we switch from one to the other at their intersection at \ensuremath{\sim}22~\ensuremath{\mathrm{R_J}}.  It is apparent that, although the estimates using the \cite{frank02a} profile is in reasonable agreement at \ensuremath{\sim}8~\ensuremath{\mathrm{R_J}} with the \cite{bagenal81a} results, the constant value taken by \cite{caudal86} in the region beyond significantly overestimates the \cite{frank02a} results over most of this region.  We thus employ a revised estimate of the number of ions per Weber beyond 8~\ensuremath{\mathrm{R_J}} given by the mean of the log of the \cite{frank02a} values, i.e.\ $10^{\langle\log N_c\rangle} = 8.1\times 10^{21}$~ ions per Weber.  A constant value has been maintained both for simplicity and since, while the flux tube volumes obtained by \cite{caudal86} are the most reasonable to use, the values have not been verified experimentally, such that here we simply use them to obtain an appropriate spot value for the number of ions per Weber.  The values of $N_c$ used in this study are therefore shown by the solid line in Figure~\ref{fig:nc}.  A second minor modification of the thermal plasma parameters concerns the plasma temperature.  As shown in equation~\ref{eq:ktc}, \cite{caudal86} used $kT_c(\rho_\circ{}) = 10(\rho_\circ{}/R_J)$~eV beyond 9~\ensuremath{\mathrm{R_J}}.  However, \cite{frank02a} report thermal plasma temperatures of \ensuremath{\sim}500~eV at \ensuremath{\sim}25~\ensuremath{\mathrm{R_J}}, such that here we instead employ $kT_c(\rho_\circ{}) = 20(\rho_\circ{}/R_J)$~eV beyond 9~\ensuremath{\mathrm{R_J}}.  Overall, therefore, on the basis of the Galileo results presented by \cite{frank02a}, the cold plasma in the model used in this study is slightly warmer and less dense than that used in \emph{Caudal's}~[1986] model.  \\

We now discuss the treatment of the plasma angular velocity.  As discussed in Section~\ref{sec:disc}, \cite{caudal86} used the fixed angular velocity profile for a dipole field given by equation~\ref{eq:hillsol} at every iteration of the model.  Here, we also initiate the model with a dipole magnetic field across the entire domain and then iteration proceeds according to his Figure 4, except that we instead solve radially outward numerically at every iteration equation~\ref{eq:hp} with $F_e$ and $|B_{ze}|$ computed from the equatorial values of $\alpha_{n-1}$ using equations~\ref{eq:alphaf} and \ref{eq:bz}, respectively, to obtain the angular velocity profile consistent with that iteration of the magnetic field, rather than simply always using the dipole solution of the Hill-Pontius equation.  The resulting angular velocity profile is then employed in equation~\ref{eq:l}, and thence the function $g_c$ given by equation~\ref{eq:gcold}.  Thus, the current sheet magnetic field and angular velocity profiles are always consistent with each other, and evolve together toward a converged self-consistent solution.  The resulting M-I currents are calculated upon convergence, defined as being when the maximum relative difference in the values of $\alpha$ between one iteration and the next is less than 0.5\%.   In some cases, the extra degree of freedom introduced by the modification of the plasma angular velocity profile between successive iterations results in the model reaching a `quasi-steady' state, rather than true convergence, in which the model perpetually fluctuates about a set of values.  In these cases the model run is stopped and the mean of the results of the last 5 iterations is used, a number which yields results representative of the set of profiles obtained when the model has reached a quasi-steady state.\\

In the results that follow we have employed ranges of values of the ionospheric Pedersen conductance and the iogenic plasma mass outflow rate, two parameters whose exact values are unknown and the effects of which on the M-I coupling current system (assuming a fixed magnetic field model) have been studied previously \citep{nichols03}.  In this study, the Pedersen conductance is assumed for simplicity to be constant, since this allows easy comparison with the analytic results of \cite{nichols03}, although we note that in reality feedback resulting from the precipitating electron flux will enhance the conductivity in the auroral region, modifying the plasma flow and current profiles as shown by \cite{nichols04}.  The iogenic mass outflow rate can be treated in two ways.  First, if the cold ion number density is assumed constant, the mass outflow rate is then related solely to the rate of outward transport of iogenic plasma, such that higher mass outflow rates equate to faster outward transport.  On the other hand, if the outward transport rate is instead assumed constant, the cold ion number density is proportional to the mass outflow rate.  Taking the canonical value of $\dot{M}=1000\;\mathrm{kg\;s^{-1}}$ as a reference, the cold ion number density values described above are then modified by

\begin{equation}
	N_c^\prime=\left(\frac{\dot{M}}{1000\;\mathrm{kg\;s^{-1}}}\right)N_c\;\;.
	\label{eq:ncprime}
\end{equation}

\noindent In Section~\ref{sec:results} we thus show results for both these scenarios, and we note that the reality will probably lie somewhere between these two cases.

\section{Results} 
\label{sec:results}

\subsection{Results with $N_c$ independent of \ensuremath{\dot{M}}}
\label{sec:resindep}

\begin{figure}
 \noindent\includegraphics[width=15pc]{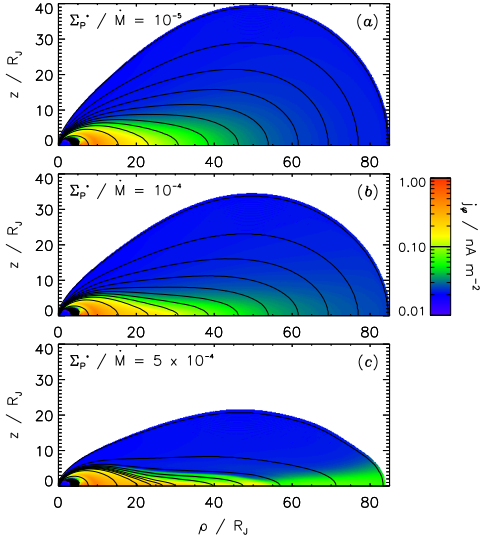}
\caption{
Plot showing the magnetic field and current sheet structures as computed using 3 values of the quotient (\ensuremath{\Sigma_P^*} / \ensuremath{\dot{M}}), i.e.\ (a) $10^{-5}$, (b) $10^{-4}$, and (c) $5 \times 10^{-4}\; \mathrm{mho\;s\;kg^{-1}}$.  The black lines are contours of $\alpha$, thus indicating magnetic field lines, and the colours indicate the azimuthal current density $j_\varphi$ in $\mathrm{nA\;m^{-2}}$.  
}
\label{fig:contours}
\end{figure}

We now present the results obtained using the model described in Section~\ref{sec:theory}.  We first show in Figure~\ref{fig:contours} the structures of the magnetic field (black contours) and total azimuthal current density (colours) as computed using 3 values for the quotient $(\Sigma_P^* /\dot{M})$~=~$10^{-5}$, $10^{-4}$, and $5 \times 10^{-4}\; \mathrm{mho\;s\;kg^{-1}}$.  These roughly bracket both the value assumed by \cite{caudal86} and the revised value of \cite{hill01} as discussed in Section~\ref{sec:theory}.  Note that here we keep the cold plasma density independent of the plasma mass outflow rate.  \cite{hill79} showed that higher values of $(\Sigma_P^* /\dot{M})$ result in higher plasma angular velocity values, and it is apparent that higher values of this quotient result in a more stretched magnetic field structure with a thinner, more intense current sheet, particularly in the region outward of \ensuremath{\sim}20~\ensuremath{\mathrm{R_J}}.  Specifically, the half-width of the current sheet in the middle magnetosphere is typically \ensuremath{\sim}8-10, \ensuremath{\sim}6-8, and \ensuremath{\sim}3-5~\ensuremath{\mathrm{R_J}} for $(\Sigma_P^* /\dot{M})$~=~$10^{-5}$, $10^{-4}$, and $5 \times 10^{-4}\; \mathrm{mho\;s\;kg^{-1}}$, respectively.  These are all somewhat larger than the value of 2.5~\ensuremath{\mathrm{R_J}} employed in the empirical `Voyager-1/Pioneer-10' (`CAN') current sheet field model of \cite{connerney81}, with the $(\Sigma_P^* /\dot{M})$~=~$5 \times 10^{-4}\; \mathrm{mho\;s\;kg^{-1}}$ result being most consistent with the latter. The increased azimuthal current for higher values of $(\Sigma_P^* /\dot{M})$ is required to balance the elevated centrifugal force imparted by the faster-rotating equatorial plasma for higher values of $(\Sigma_P^* /\dot{M})$. \\

\begin{figure}
 \noindent\includegraphics[width=13pc]{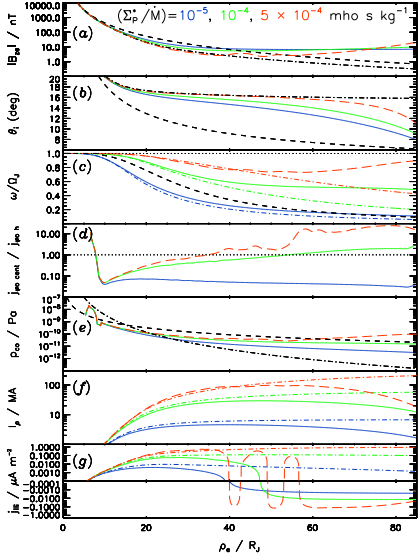}
\caption{
\figsquash Plot showing magnetodisc and current system parameters computed for $(\Sigma_P^* /\dot{M})$~=~$10^{-5}$ (blue), $10^{-4}$ (green), and $5 \times 10^{-4}$ (red) $\; \mathrm{mho\;s\;kg^{-1}}$, plotted versus equatorial radial distance.  Parameters shown are (a) the magnitude of the north-south magnetic field threading the equatorial plane \ensuremath{|B_{ze}|} in nT, (b) the ionospheric co-latitude to which the magnetic field maps $\theta_i$ in degrees, (c) the equatorial plasma angular velocity normalised to the planet's rotation rate $(\omega/\Omega_J)$, (d) the ratio of the equatorial azimuthal current density associated with the cold plasma centrifugal force to that of the hot plasma pressure $(j_{\phi\,\circ\;\mathrm{cent}} / j_{\phi\,\circ\;\mathrm{h}})$, (e) the cold plasma pressure $P_{c\,\circ}$ in Pa, (f) the azimuthally-integrated equatorial radial current $I_\rho$ in MA, and (g) the field-aligned current density at the top of the ionosphere \ensuremath{j_{\|i}} in \ensuremath{\mu \mathrm{A\;m^{-2}}}.  In Figures~\ref{fig:eqsig}a and \ref{fig:eqsig}b the dashed black lines show the planetary dipole values, and the dot-dashed black lines show the values using the model of \cite{nichols04}.  In Figure~\ref{fig:eqsig}c the horizontal dotted line indicates rigid corotation and the dashed black line indicates the profile taken by \cite{caudal86}.  In Figure~\ref{fig:eqsig}d the horizontal dotted line indicates where the azimuthal current densities associated with the cold plasma centrifugal force and the hot plasma pressure are equal.  In Figure~\ref{fig:eqsig}e the dashed and dot-dashed black lines show the power laws given by \cite{frank02a}. Note that both the positive and negative values in Figure~\ref{fig:eqsig}g are plotted on logarithmic scales, such that the horizontal line is at $\pm0.0001$, and the resulting apparent discontinuities at the transition points are simply artefacts of the plotting scale.  In Figures~\ref{fig:eqsig}f and \ref{fig:eqsig}g the canonical value of \ensuremath{\dot{M}}~=~1000~\ensuremath{\mathrm{kg\;s^{-1}}} is used.  In all panels the solid coloured lines indicate results from model runs which converged, while long-dashed lines indicate results from model runs which have reached a `quasi-steady' state as discussed in Section~\ref{sec:application}.   In Figures~\ref{fig:eqsig}c, \ref{fig:eqsig}f and \ref{fig:eqsig}g the coloured dot-dashed lines indicate results obtained using the fixed magnetic field model of \cite{nichols04} for comparison.  
}
\label{fig:eqsig}
\end{figure}

This can be further appreciated from Figure~\ref{fig:eqsig}, in which we show various parameters associated with the magnetodisc model and M-I coupling current system for each of the above values of $(\Sigma_P^* /\dot{M})$, and where for the purposes of the M-I coupling current calculations we take the canonical value of $\dot{M}=1000\;\mathrm{kg\;s^{-1}}$.  Specifically, we show, from top to bottom, the magnitude of the north-south magnetic field threading the equatorial plane \ensuremath{|B_{ze}|} in nT, the ionospheric co-latitude to which the magnetic field maps $\theta_i$  in degrees, the equatorial plasma angular velocity normalised to the planet's rotation rate $(\omega/\Omega_J)$, the ratio of the equatorial azimuthal current density associated with the cold plasma centrifugal force to that of the hot plasma pressure  $(j_{\phi\,\circ\;\mathrm{cent}} / j_{\phi\,\circ\;\mathrm{h}})$, the cold plasma pressure $P_{c\,\circ}$ in Pa, the azimuthally-integrated equatorial radial current $I_\rho$ in MA, and finally the field-aligned current density \ensuremath{j_{\|i}} at the top of the ionosphere in \ensuremath{\mu \mathrm{A\;m^{-2}}}, all versus equatorial radial distance in \ensuremath{\mathrm{R_J}}.  Note that the solid coloured lines indicate results from model runs which converged, while long-dashed lines indicate results from model runs which have reached a `quasi-steady' state as discussed in Section~\ref{sec:application}.  Starting with the equatorial magnetic field strength \ensuremath{|B_{ze}|} shown in Figure~\ref{fig:eqsig}a, it is first evident that all three model results are similar out to \ensuremath{\sim}15~\ensuremath{\mathrm{R_J}}, beyond which they diverge.  Also shown in Figure~\ref{fig:eqsig}a for comparison are the magnetic field strengths given by the pure planetary dipole (dashed black line), given by

\begin{equation}
	B_{ze\,\mathrm{dip}}(\rho_e) = -B_J \left(\frac{R_J}{\ensuremath{\rho_e}}\right)^3\;\;
	\label{eq:bzedip}
\end{equation}

\noindent and the `CAN-KK' current sheet magnetic field model of \cite{nichols04} (dot-dashed line), given by

\begin{align}
	&B_{ze\,\mathrm{CAN-KK}}(\rho_e)= \nonumber \\
	&-\left\{B_\circ^\prime\left(\frac{R_J}{\rho_e}\right)^3\exp\left[-\left(\frac{\rho_e}{\rho_e^*}\right)^{5/2}\right]+B_\circ\left(\frac{R_J}{\rho_e}\right)^m\right\},
	\label{eq:bze}
\end{align}

\noindent where $B_\circ^\prime=3.335\times10^5$~nT, $\rho_e^*=14.501$~\ensuremath{\mathrm{R_p}}, $B_\circ=5.4\times10^4$~nT, and $m=2.71$.  This form closely approximates the field model used by \cite{cowley01} and \cite{cowley02,cowley03a}, who employed the CAN field model of \cite{connerney81} in the inner region and the Voyager-1 (`KK') outbound pass model of \cite{khurana93} in the outer region.  In the model results obtained in this study, the \ensuremath{|B_{ze}|} values are less than those for the dipole in the inner region owing to the radial distention of the field by the current sheet.  All three results are reasonably consistent with the CAN-KK model to distances of \ensuremath{\sim}20~\ensuremath{\mathrm{R_J}}, but $(\Sigma_P^* /\dot{M})$~=~$5 \times 10^{-4}\; \mathrm{mho\;s\;kg^{-1}}$ again gives the best agreement, roughly tracking the CAN-KK model values out to \ensuremath{\sim}40~\ensuremath{\mathrm{R_J}}.  We note that the slight jitter in the latter \ensuremath{|B_{ze}|} profile between \ensuremath{\sim}40 and \ensuremath{\sim}60~\ensuremath{\mathrm{R_J}} is representative of the spontaneous instability which prohibits models runs with higher values of $(\Sigma_P^* /\dot{M})$ from truly converging, and which, as mentioned by \cite{caudal86}, can lead to the formation of neutral points for more stretched magnetodiscs.  In the outer region, the \ensuremath{|B_{ze}|} values become greater than those for the dipole, with the transitions occurring at \ensuremath{\sim}38, \ensuremath{\sim}42, and \ensuremath{\sim}52~\ensuremath{\mathrm{R_J}} for $(\Sigma_P^* /\dot{M})$~=~$10^{-5}$, $10^{-4}$, and $5 \times 10^{-4}\; \mathrm{mho\;s\;kg^{-1}}$, respectively.  This transition, also originally noted by \cite{caudal86}, is due to the outer fringing fields of the current sheet, and the outer-most values of \ensuremath{|B_{ze}|} of \ensuremath{\sim}10-20~nT are consistent with the values of \ensuremath{\sim}16~nT observed just inside the magnetopause \citep{acuna83a}.  Note that in contrast, in the radial range of Figure~\ref{fig:eqsig} the `CAN-KK' values are always less than those of the dipole, indicating that the current sheet in the pre-dawn region of the Voyager-1 outbound pass was evidently extended due to the distant (\ensuremath{\sim}160~\ensuremath{\mathrm{R_J}}) magnetopause in this region \citep{acuna83a}.  \\

Figure~\ref{fig:eqsig}b shows the ionospheric co-latitude to which the magnetic field maps, calculated using equations~\ref{eq:fi} and \ref{eq:alphaf}.  Also shown are the values for the planetary dipole  (dashed black line), for which 

\begin{equation}
	F_{e\,\mathrm{dip}}(\rho_e) = \frac{B_JR_J^3}{\ensuremath{\rho_e}}\;\;,
	\label{eq:fedip}
\end{equation}

\noindent and the CAN-KK field model (dot-dashed line), for which

\begin{align}
	&F_{e\,\mathrm{CAN-KK}}(\rho_e) = \nonumber \\
	&F_\infty+\frac{B_\circ^\prime R_J^3}{2.5\rho_e^*}\Gamma\left[-\frac{2}{5},\left(\frac{\rho_e}{\rho_e^*}\right)^{5/2}\right]+ 
	\frac{B_\circ}{(m-2)}\left(\frac{R_J}{\rho_e}\right)^{m-2}\;\;,
	\label{eq:fecankk}
\end{align}

\noindent where $F_\infty\approx 2.841\times 10^4$~nT~\ensuremath{\mathrm{R_J}^2} is the value at infinity, and $\Gamma(a,z)=\int_z^\infty t^{a-1}e^{-t}\:dt$ is the incomplete gamma function.  It is apparent that for each value of $(\Sigma_P^* /\dot{M})$ used the field line mapping is more consistent with that of the CAN-KK field model than the dipole, although the elevated values of \ensuremath{|B_{ze}|} in the outer region relative to the CAN-KK values results in a broadening of the ionospheric latitudinal band to which the outer magnetosphere maps. It is evident, however, that for increased values of $(\Sigma_P^* /\dot{M})$ the middle magnetosphere field lines map to a modestly more equatorward and thinner latitudinal band in the ionosphere.  For example, field lines threading the equatorial plane between 20-60~\ensuremath{\mathrm{R_J}} map to between \ensuremath{\sim}12.6-16.6$^\circ$, \ensuremath{\sim}14.0-16.9$^\circ$, \ensuremath{\sim}15.6-17.1$^\circ$ for $(\Sigma_P^* /\dot{M})$~=~$10^{-5}$, $10^{-4}$, and $5 \times 10^{-4}\; \mathrm{mho\;s\;kg^{-1}}$, respectively, with the latter result being most consistent with the CAN-KK model.  In addition, it is worth noting that in this model the ionospheric co-latitudes of the last closed field line are \ensuremath{\sim}7.9$^\circ$, \ensuremath{\sim}9.0$^\circ$, and \ensuremath{\sim}11.0$^\circ$ for $(\Sigma_P^* /\dot{M})$~=~$10^{-5}$, $10^{-4}$, and $5 \times 10^{-4}\; \mathrm{mho\;s\;kg^{-1}}$, respectively.  The latter value is in excellent agreement with the value of \ensuremath{\sim}11$^\circ$ recently determined by \cite{vogt11a}, and is also consistent with the value of 10.25\ensuremath{^\circ{}} used by \cite{cowley05a} in their global model of Jupiter's polar ionospheric flows.\\

The angular velocity of the equatorial plasma as computed by the present model is shown in Figure~\ref{fig:eqsig}c, along with the angular velocity profiles calculated using the fixed CAN-KK field model and the profile assumed by \cite{caudal86} for comparison.  In all cases the profiles calculated here are similar to those obtained using the CAN-KK field model in the inner region, and deviate toward higher values in the outer region due to the increased $\mathbf{j}\times\mathbf{B}$ force owing to the elevated values of \ensuremath{|B_{ze}|} relative to the CAN-KK model.  As mentioned in Section~\ref{sec:theory}, the angular velocity profile used by \cite{caudal86} is roughly consistent with the lowest value of $(\Sigma_P^* /\dot{M})$ used here, and reduces to \ensuremath{\sim}0.17 by 60~\ensuremath{\mathrm{R_J}}.  On the other hand, observational studies such as \cite{kane95} have reported that the plasma angular velocities remain at \ensuremath{\sim}0.5 out to \ensuremath{\sim}60~\ensuremath{\mathrm{R_J}}.  The present angular velocity profile which best fits this behaviour is that produced using $(\Sigma_P^* /\dot{M})=10^{-4}\;\mathrm{mho\;s\;kg^{-1}}$, which is also generally consistent with the values at the plasma sheet crossings obtained by \cite{mcnutt81a} (i.e.\ those values at the local maxima in the Voyager 1 data shown in their Figure~21).  Lower and higher values of $(\Sigma_P^* /\dot{M})$ then produce angular velocity profiles which are overall respectively somewhat lower and higher than observations suggest.  \\

The effect of the plasma angular velocity on the azimuthal current is shown in Figure~\ref{fig:eqsig}d, in which we plot the ratio of the equatorial azimuthal current density associated with the centrifugal force to that of the hot plasma pressure $(j_{\phi\,\circ\;\mathrm{cent}} / j_{\phi\,\circ\;\mathrm{h}})$, giving an indication as to which of these two components of the azimuthal current is dominant.  \cite{caudal86} concluded that the latitude-integrated current associated with the hot plasma pressure dominates both the cold plasma pressure current and the centrifugal force current over the whole of the magnetosphere.  This result was supported by \cite{achilleos10a}, although these authors also showed that in the original \cite{caudal86} model the effect of the centrifugal force strongly peaks near \ensuremath{\sim}27~\ensuremath{\mathrm{R_J}}, such that equatorial current densities associated with the hot plasma pressure and centrifugal force become comparable between \ensuremath{\sim}20-30~\ensuremath{\mathrm{R_J}}, a concern which was originally raised by \cite{mauk87a} on the basis that it apparently contradicts observation \citep{mcnutt83a,mcnutt84a}.  It is therefore worth noting that the revised cold plasma input parameters employed in our model eliminate this effect here, and considering first the current ratio profile for $(\Sigma_P^* /\dot{M})$~=~$10^{-5}$ profile, it is apparent that the hot plasma current is significantly larger than that of the centrifugal force over essentially all the magnetosphere.  However, this is not the case for the higher values of $(\Sigma_P^* /\dot{M})$, for which the centrifugal force current exceeds the hot plasma pressure current outward of \ensuremath{\sim}54 and \ensuremath{\sim}34~\ensuremath{\mathrm{R_J}} for $(\Sigma_P^* /\dot{M})$~=~$10^{-4}$, and $5 \times 10^{-4}\; \mathrm{mho\;s\;kg^{-1}}$, respectively. It should be noted that this does not contradict the conclusions of \cite{mcnutt83a,mcnutt84a} and \cite{mauk87a}, which were based on Voyager data that were obtained at current sheet crossings within 40~\ensuremath{\mathrm{R_J}} and that are somewhat sparse beyond \ensuremath{\sim}30~\ensuremath{\mathrm{R_J}} (see, e.g.\ Figure~2 of \cite{mcnutt83a}).  \cite{mcnutt84a} and \cite{mauk87a} computed the ratio of the rotational kinetic energy density to magnetic energy density, which can be thought of a `plasma beta for bulk rotation', comparable to the traditional plasma beta  $\beta=(P/P_B)$, where $P_B=B^2/2\mu_\circ$ is the magnetic energy density (note they termed this quantity $M^2$, since it is equal to the square of the Alfv\'enic Mach number).  \cite{achilleos10a} pointed out that the plasma beta for bulk rotation is given by $\beta_\mathrm{cent}=(\beta_c\rho^2/2l^2)$, and thus confirmed that in \emph{Caudal's}~[1986] model the hot plasma beta $\beta_h$ dominates the bulk rotation beta $\beta_\mathrm{cent}$ beyond \ensuremath{\sim}40~\ensuremath{\mathrm{R_J}}.  We have calculated the ratio $(\beta_\mathrm{cent}/\beta_h)$ using our model results and, while for clarity we have not plotted the profiles in Figure~\ref{fig:eqsig}, we note that they are very similar to those for $(j_{\phi\,\circ\;\mathrm{cent}} / j_{\phi\,\circ\;\mathrm{h}})$. \cite{achilleos10a} showed that in the original model of \cite{caudal86}, $\beta_\mathrm{cent}$ peaks near \ensuremath{\sim}25~\ensuremath{\mathrm{R_J}} at \ensuremath{\sim}16, whereas \cite{mcnutt84a} obtained values of \ensuremath{\sim}3 near 25~\ensuremath{\mathrm{R_J}}.  In our results $\beta_\mathrm{cent}$~=~0.75, 6.63, and 24.96 at 25~\ensuremath{\mathrm{R_J}} for  $(\Sigma_P^* /\dot{M})$~=~$10^{-5}$, $10^{-4}$, and $5 \times 10^{-4}\; \mathrm{mho\;s\;kg^{-1}}$, respectively, with the value for $(\Sigma_P^* /\dot{M})~=~10^{-4}\;\mathrm{mho\;s\;kg^{-1}}$ thus being in most agreement with observations. \\

The cold plasma pressure computed in this model is plotted in Figure~\ref{fig:eqsig}e, along with power laws fitted by \cite{frank02a} to the pressure values as measured by the Galileo spacecraft.  These fits are given by

\begin{equation}
	P_{c\,\circ}(\rho_\circ{}) = 
	  \begin{cases}
		  1.9 \times{} 10^{-4}\; \rho_\circ{} ^{-4.71}\; \mathrm{ Pa} & \text{if } \rho_\circ{} < 20\;\mathrm{R_J} \\
		  8.6 \times{} 10^{-8}\; \rho_\circ{} ^{-1.87}\; \mathrm{ Pa} & \text{if }  \rho_\circ{} > 50\;\mathrm{R_J}  \;\;,
	  \end{cases}
	\label{eq:fpc}
\end{equation}

\noindent shown by the dot-dashed and dashed black lines, respectively, although we note that in Figure~8 of \cite{frank02a}, the scatter in the measured values is generally at least an order of magnitude, and in the region $20<(\rho_\circ/R_J)<50$ the points generally lie between the two power laws.    All three cold plasma pressure profiles are similar out to distances of \ensuremath{\sim}15~\ensuremath{\mathrm{R_J}}, beyond which the profiles for $(\Sigma_P^* /\dot{M})$~=~$10^{-4}$ and $5 \times 10^{-4}\; \mathrm{mho\;s\;kg^{-1}}$ are in best agreement with the observed profile.  \\

Figure~\ref{fig:eqsig}f shows the azimuth-integrated equatorial radial current computed from the plasma angular velocity and magnetic field profiles using equation~\ref{eq:totip}, where we note that for the M-I coupling current equations we explicitly take the canonical value of \ensuremath{\dot{M}}~=~1000~\ensuremath{\mathrm{kg\;s^{-1}}}, such that \ensuremath{\Sigma_P^*}~=~0.01, 0.1, and 0.5~mho.  The solid lines show the results obtained using the magnetic field model discussed here, while the dot-dashed lines show the profiles obtained using the empirical CAN-KK magnetic field model employed in previous studies for comparison.  It is evident that the current profiles are similar to the results for the CAN-KK model out to \ensuremath{\sim}40-50~\ensuremath{\mathrm{R_J}}, beyond which they reduce to smaller values owing to the lower values of $F_e$ in the outer region relative to the CAN-KK values due to the current sheet outer fringing field.  \cite{nichols04} used the midnight sector Galileo $B_\varphi$ data of \cite{khurana01} to show that the observed values of $I_\rho$ increase rapidly in the inner region, between \ensuremath{\sim}15 and 25~\ensuremath{\mathrm{R_J}}, before plateauing at \ensuremath{\sim}100~MA at distances beyond, out to \ensuremath{\sim}100~\ensuremath{\mathrm{R_J}} (see, e.g.\ their Figure~12).  It is worth noting that the \cite{khurana01} data obtained at midnight is unconstrained by an assumed magnetopause distance of 85~\ensuremath{\mathrm{R_J}}, such that it is not surprising that the decrease in the outer region is not evident in those data.  This aside, the current profile which best fits this pattern is that for \ensuremath{\Sigma_P^*}~=~0.5~mho.\\

The resulting field-aligned current at the top of the ionosphere computed using equation~\ref{eq:jpari} is then plotted in Figure~\ref{fig:eqsig}g.  In the inner region the currents are upward and peak at similar values to those obtained using the CAN-KK field model, at radial distances of \ensuremath{\sim}22, 28, and 33~\ensuremath{\mathrm{R_J}} for \ensuremath{\Sigma_P^*}~=~0.01, 0.1, and 0.5~mho, respectively.  However, the decreasing values of $I_\rho$ in the outer region result in a reversal of the field-aligned current at \ensuremath{\sim}40-60~\ensuremath{\mathrm{R_J}}, such that the current is then downward in the region beyond.  Note that the oscillation in the \ensuremath{\Sigma_P^*}~=~0.5~mho profile between \ensuremath{\sim}40-60~\ensuremath{\mathrm{R_J}} is due to the instability in the magnetic field model as discussed above.  While we note that such layering of upward and downward field-aligned current has been observed in Jupiter's middle magnetosphere by \cite{mauk07a}, we do not wish to infer too much from the structure in our results, and simply note that the overall structure is that of consistent upward current inward of \ensuremath{\sim}40~\ensuremath{\mathrm{R_J}} and downward current outward of \ensuremath{\sim}60~\ensuremath{\mathrm{R_J}}. This confinement of the upward field-aligned current to the region inward of \ensuremath{\sim}40-60~\ensuremath{\mathrm{R_J}}, depending on $(\Sigma_P^* /\dot{M})$,  is consistent with the results of \cite{vogt11a}, who showed using flux equivalence calculations that the poleward boundary of the main auroral oval maps to \ensuremath{\sim}30-60~\ensuremath{\mathrm{R_J}} depending on local time, and we also note that \cite{khurana01} showed using Galileo data that the main oval field-aligned currents flow inward of 30~\ensuremath{\mathrm{R_J}}.  Inclusion of local time asymmetry is not possible in our axisymmetric model, but the overall results are broadly consistent with the observations of \cite{khurana01} and \cite{vogt11a}.  The downward current in the region outward of \ensuremath{\sim}40-60~\ensuremath{\mathrm{R_J}} thus corresponds to the dark polar region just poleward of main oval, which typically exists on the dawn side but sometimes extends to all local times \citep{grodent03a, nichols09b}.  Note that while \cite{nichols04} showed that the modulation of the ionospheric Pedersen conductivity by auroral electron precipitation concentrates the peak field-aligned current in the \ensuremath{\sim}20-40~\ensuremath{\mathrm{R_J}} region, in their model the field-aligned current was still upward throughout the magnetosphere, albeit at low values in the outer region.  In addition, while \cite{cowley05a} included a region of downward current in the outer magnetosphere by design of their specified plasma velocity profiles, the results presented here are the first to self-consistently produce this downward current region.  The latter authors also showed that a second sheet of upward field-aligned current should exist, associated with the ionospheric flow shear at the boundary between open and closed field lines, and indeed it is thought that Saturn's main auroral oval is due to such a layer between the outer edge of the ring current and the open-closed field line boundary \citep{badman06a,bunce08a}. Since our model only includes closed field lines, we do not consider this second layer of upward current and simply note that it will act to modify the field-aligned current profiles in the very outer region from those computed here.  \\

\begin{figure}
 \noindent\includegraphics[width=19pc]{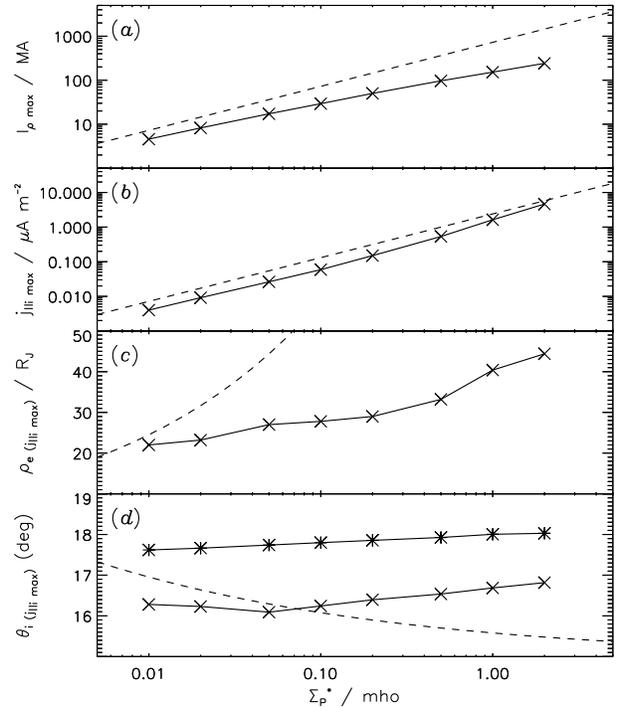}
\caption{
\figsquash Plot showing (a) the maximum azimuth-integrated radial current, (b) the maximum field-aligned current density at the top of the ionosphere, (c) the equatorial distance of the maximum field-aligned current, and (d) the ionospheric co-latitudes of the maximum field-aligned current (joined crosses) and Ganymede footprint (joined asterisks), all versus ionospheric Pedersen conductance.  Also shown by the dashed lines are the results for a power law current sheet magnetic field.  
}
\label{fig:maxsig}
\end{figure}

Overall, then, it is apparent that the results for $(\Sigma_P^* /\dot{M})$~=~$10^{-4}$ and $5 \times 10^{-4}\; \mathrm{mho\;s\;kg^{-1}}$ provide the best agreement with various sets of observations, with the magnetic field (both \ensuremath{|B_{ze}|} and $B_\varphi$) most consistent with the latter, and plasma data (angular velocity and pressure) in best agreement with the former.  It is, however, instructive to examine how the M-I coupling currents vary in peak magnitude and location over a range of values of $(\Sigma_P^* /\dot{M})$, which we thus show in Figure~\ref{fig:maxsig}, again taking here \ensuremath{\dot{M}}~=~1000~\ensuremath{\mathrm{kg\;s^{-1}}} (note that we consider the effect of changing \ensuremath{\dot{M}} in Section~\ref{sec:resdep}). From top to bottom, the joined crosses in Figure~\ref{fig:maxsig} show the maximum azimuth-integrated equatorial radial current, the maximum upward field-aligned current density at the top of the ionosphere, the equatorial radial distance of the peak upward field-aligned current, and finally the ionospheric co-latitude of the peak upward field-aligned current.  Also shown for comparison by the dashed lines in Figure~\ref{fig:maxsig} are results calculated using the analytical solution of the Hill-Pontius equation (equation~\ref{eq:hp}) obtained by \cite{nichols03} for a power law current sheet magnetic field which maps to a thin latitude band in the ionosphere, such as that given by the second term in equation~\ref{eq:bze}, thus appropriate for the jovian middle magnetosphere.  The analytic power law field result for the maximum azimuth-integrated radial current is given by

\begin{equation}
	I_{\rho\,\mathrm{max}} = 8\pi \ensuremath{\Sigma_P^*} \ensuremath{\Omega_J} F_\circ\;\;,
	\label{eq:maxirho}
\end{equation}

\noindent where $F_\circ$ is the value of $F_e$ at the location of the latitude band, taken by \cite{nichols03} to be $F_\circ = F_e(70\;R_J) \simeq 3.22 \times 10^4 \;\mathrm{nT\;R_J^2} $, a representative value for the middle magnetosphere current sheet.  The maximum value for the power law field strictly occurs at \ensuremath{\rho_e=\infty}, while at large but finite distances in the numeric solution using the full empirical field model (e.g.\ beyond 1000~\ensuremath{\mathrm{R_J}} for \ensuremath{\Sigma_P^*}~=~0.5 mho).  It is apparent that the maximum radial current computed using the model employed here increases less quickly with \ensuremath{\Sigma_P^*} than for the power law field, i.e.\ from \ensuremath{\sim}5~MA at \ensuremath{\Sigma_P^*}~=~0.01~mho to \ensuremath{\sim}242~MA at \ensuremath{\Sigma_P^*}~=~2~mho.  This occurs since, for a power law current sheet field the total azimuthal current increases monotonically toward the maximum value given by equation~\ref{eq:maxirho} at a rate determined solely by the corotation breakdown distance, given by $\rho_H$ for a dipole field and by

\begin{equation}
	\left(\frac{\rho_{H\mathrm{\,cs}}}{R_J}\right) = \left(\frac{2\pi \ensuremath{\Sigma_P^*}B_\circ F_\circ}{\ensuremath{\dot{M}}}\right)^{1/m}
	\label{eq:rhohcs}
\end{equation}

\noindent for the power law current sheet field.  The radial current profiles obtained using the model presented here, however, are also constrained by the assumed magnetopause distance and thus drop away from the power law profiles at distances increasingly small relative to $\rho_{H\mathrm{\,cs}}$ as $(\Sigma_P^* /\dot{M})$ increases, such that the peak current rises less quickly with $\Sigma_P^*$ than for the power law field.  \\

The maximum field-aligned current current density at the top of the ionosphere is plotted in Figure~\ref{fig:maxsig}b, alongside the result for the power law current sheet field, shown by \cite{nichols03} to be

\begin{equation}
	\ensuremath{j_{\|i\,\mathrm{max}}} \simeq 3.05 
	\left(\frac{F_\circ}{B_\circ R_J^2}\right)\left(\frac{\rho_{H\mathrm{\,cs}}}{R_J}\right)^{m-2}
	\ensuremath{\Sigma_P^*}B_J \ensuremath{\Omega_J}\;\;.
	\label{eq:maxjpi}
\end{equation}

\noindent It is evident that the maximum field-aligned current density computed here increases with \ensuremath{\Sigma_P^*} similarly as does the result for power law field, i.e.\ from \ensuremath{\sim}0.004~\ensuremath{\mu \mathrm{A\;m^{-2}}} at \ensuremath{\Sigma_P^*}~=~0.01~mho to \ensuremath{\sim}4.6~\ensuremath{\mu \mathrm{A\;m^{-2}}} at \ensuremath{\Sigma_P^*}~=~2~mho, such that the latter is a reasonable approximation for the results obtained here.  \\

The same is not true, however, for the equatorial radial distance of the peak field-aligned current density, shown in Figure~\ref{fig:maxsig}c, in which the distance for the power law field, given by

\begin{equation}
	\left(\frac{\rho_{e\,(j_{\|i\,\mathrm{max}})}}{R_J}\right) \simeq 2.38 \left(\frac{\rho_{H\mathrm{\,cs}}}{R_J}\right)\;\;,
	\label{eq:maxrhoe}
\end{equation}

\noindent rises much more quickly than do the results here, which increase from 22~\ensuremath{\mathrm{R_J}} at \ensuremath{\Sigma_P^*}~=~0.01~mho to \ensuremath{\sim}44~\ensuremath{\mathrm{R_J}} at \ensuremath{\Sigma_P^*}~=~2~mho.  This is again due to constraint by the finite magnetopause distance in the current sheet field model employed here, rather than the power law field which simply decreases monotonically toward  \ensuremath{\rho_e=\infty}.  The ionospheric co-latitudes of these peak field-aligned current locations are shown in Figure~\ref{fig:maxsig}d by the joined crosses, along with the mapped location of Ganymede's orbit at 15~\ensuremath{\mathrm{R_J}}, shown by the joined asterisks. Again, shown by the dashed line for comparison is the location of the peak field-aligned current for the power law field, given by

\begin{align}
	&\theta_{i\,(j_{\|i\,\mathrm{max}})}= \nonumber \\
	&\sin^{-1}\sqrt{\frac{F_\infty}{B_JR_J^2}+
	\left[\frac{B_\circ}{(m-2)B_J}\right]\left[\frac{\rho_{e\,(j_{\|i\,\mathrm{max}})}}{R_J}\right]^{2-m}}
	\label{eq:maxthetai}\;\;,
\end{align}

\noindent which indicates that in this case the peak current shifts poleward as the equatorial radial distance increases, although, as shown in Figure~\ref{fig:eqsig}b, the co-latitude is only weakly dependent on the radial distance due to the stretching of the middle magnetosphere field lines.  However, although the equatorial radial distance of the peak field-aligned current in the results presented here increases with \ensuremath{\Sigma_P^*}, above \ensuremath{\Sigma_P^* = 0.05}~mho, the ionospheric co-latitude actually increases slowly with \ensuremath{\Sigma_P^*}, moving from \ensuremath{\sim}16.3\ensuremath{^\circ{}} for \ensuremath{\Sigma_P^*}~=~0.01~mho to \ensuremath{\sim}16.8\ensuremath{^\circ{}} for \ensuremath{\Sigma_P^*}~=~2~mho.  This arises since the outward movement of peak field-aligned current with increasing \ensuremath{\Sigma_P^*} is offset by the modified mapping of the increasingly stretched magnetic field.  This can be appreciated by examination of Figures~\ref{fig:eqsig}b and \ref{fig:eqsig}g, in which the peak field-aligned current moves outward for the blue, green and red profiles, respectively, while the associated ionospheric mapping profiles also move equatorward, counteracting the outward shift.  Considering now the co-latitude of the Ganymede footprint, it is evident that this is also only very weakly dependent on \ensuremath{\Sigma_P^*}, moving from \ensuremath{\sim}17.6\ensuremath{^\circ{}} for \ensuremath{\Sigma_P^*}~=~0.01~mho to \ensuremath{\sim}18.0\ensuremath{^\circ{}} for \ensuremath{\Sigma_P^*}~=~2~mho.  This is simply due to the fact that in these runs the magnetic field model is relatively insensitive to changes inside \ensuremath{\sim}15~\ensuremath{\mathrm{R_J}}, and, as can be seen from Figure~\ref{fig:eqsig}b, the mapped ionospheric co-latitudes of \ensuremath{\rho_e=15}~\ensuremath{\mathrm{R_J}} are very similar for all values of \ensuremath{\Sigma_P^*}. Thus, on the basis of these results, it is unlikely that a change of ionospheric conductance is responsible for the \ensuremath{\sim}3\ensuremath{^\circ{}} and \ensuremath{\sim}2\ensuremath{^\circ{}} shifts in latitude of the main oval and Ganymede footprint, respectively, reported by \cite{grodent08}.  In the following section we therefore examine the effect of changing cold plasma number density.  

\subsection{Comparison with results taking $N_c$ proportional to \ensuremath{\dot{M}}}
\label{sec:resdep}

\begin{figure}
 \noindent\includegraphics[width=19pc]{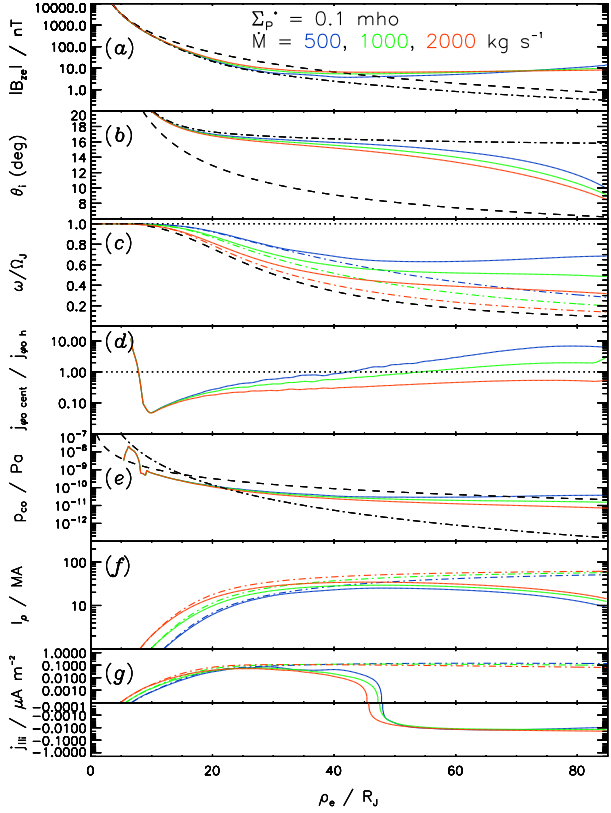}
\caption{
As for Figure~\ref{fig:eqsig}, except with \ensuremath{\Sigma_P^*}~=~0.1~mho and \ensuremath{\dot{M}}~=~500 (blue), 1000 (green), and 2000 (red)~\ensuremath{\mathrm{kg\;s^{-1}}}.    
}
\label{fig:eqmdot}
\end{figure}

\begin{figure}
 \noindent\includegraphics[width=19pc]{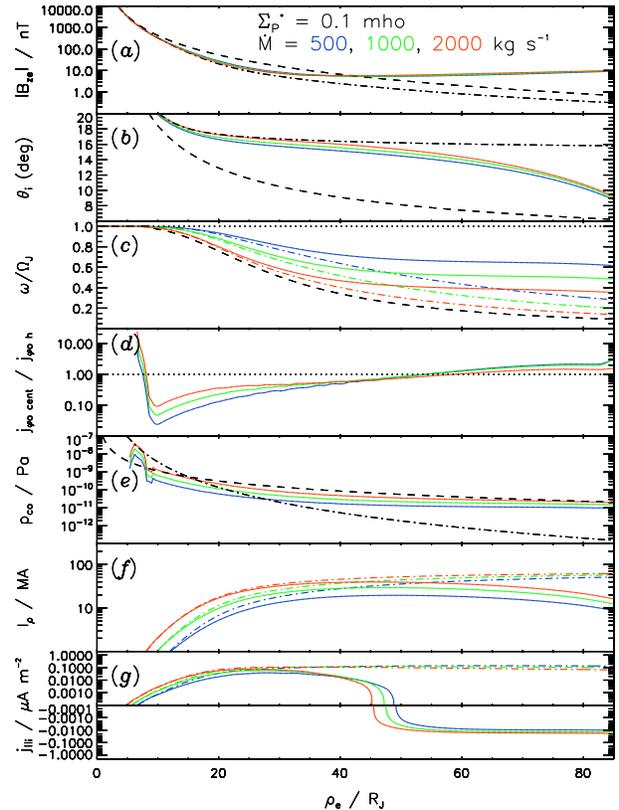}
\caption{
As for Figure~\ref{fig:eqmdot}, except with the cold plasma number density given by equation~\ref{eq:ncprime}.    
}
\label{fig:eqmod}
\end{figure}

We now compare results for which the cold plasma density is assumed constant, such that the outward transport rate is proportional to \ensuremath{\dot{M}}, with results for which the cold plasma density is taken to be given by equation~\ref{eq:ncprime}, such that in this case the outward transport rate is assumed constant.  Here, we take \ensuremath{\Sigma_P^*=0.1}~mho, and \ensuremath{\dot{M}}~=~500, 1000, and 2000~\ensuremath{\mathrm{kg\;s^{-1}}}, typical of the range of values determined by various studies \citep[e.g.][]{hill80,khurana93,delamere03a}.  Figure~\ref{fig:eqmdot} thus shows the magnetodisc and M-I coupling current system parameters for constant plasma density in the same format as for Figure~\ref{fig:eqsig}, while Figure~\ref{fig:eqmod} shows the results taking the cold plasma density to be given by equation~\ref{eq:ncprime}.  It is first evident from Figures~\ref{fig:eqmdot}a and \ref{fig:eqmod}a that taking $N_c \propto \ensuremath{\dot{M}}$ acts to suppress the divergence of the \ensuremath{|B_{ze}|} profiles in the middle magnetosphere beyond \ensuremath{\sim}20~\ensuremath{\mathrm{R_J}}.  It is, however, just apparent that for the case with $N_c \propto \ensuremath{\dot{M}}$, the higher value of \ensuremath{\dot{M}} leads to slightly lower equatorial magnetic field strengths in the region inside \ensuremath{\sim}40~\ensuremath{\mathrm{R_J}}, i.e.\ the opposite behaviour to the case with constant $N_c$.  This is more evident in Figures~\ref{fig:eqmdot}b and \ref{fig:eqmod}b, in the former of which the field maps to lower co-latitudes for higher mass outflow rates, indicating a less stretched field, while in the latter case the field maps to higher co-latitudes, indicating a more stretched field.  It is also worth noting that  for the case in \ref{fig:eqmod}b the difference in field mapping is larger at all radial distances than for \ref{fig:eqmdot}b, in which the divergence is only significant outward of \ensuremath{\sim}15~\ensuremath{\mathrm{R_J}}.  This indicates the nature of the centrifugal force acting on the plasma in the two cases, which we now discuss.\\

Figures~\ref{fig:eqmdot}c and \ref{fig:eqmod}c show that the plasma angular velocities in the two cases are very similar, with perhaps modestly increased values in Figure~\ref{fig:eqmdot}c over those in Figure~\ref{fig:eqmod}c.  The plasma angular velocity is related to the centrifugal force acting on the rotating plasma, which we recall from equation~\ref{eq:mom} is proportional to $d \omega^2$, such that if these two parameters are dependent on \ensuremath{\dot{M}}, the centrifugal force is proportional to some power of \ensuremath{\dot{M}}, i.e.\ $\ensuremath{\dot{M}}^\gamma$. An understanding of the difference in behaviour between the two cases can then be obtained if we consider the power law magnetic field approximations of \cite{nichols03}.  In this approximation, the plasma angular velocity scales with the current sheet `Hill-distance' $\rho_{H\mathrm{\,cs}}$ given by equation~\ref{eq:rhohcs}, such that $\omega \propto \ensuremath{\dot{M}}^{-1/m}$.  Hence, if the plasma density is independent of the mass outflow rate, we have $\gamma = -2/m$, while if the plasma density is proportional to the mass outflow rate we have $\gamma = 1-2/m$.  Therefore, in the former case $\gamma<0$ for all positive values of $m$ (i.e.\ for fields which decrease in magnitude with distance), such that  the centrifugal force decreases with increasing \ensuremath{\dot{M}}.  On the other hand, for the latter case we have $\gamma<0$ for $m<2$, such that the centrifugal force decreases with increasing \ensuremath{\dot{M}}, and $0<\gamma<1$ for $m>2$, such that the centrifugal force increases with increasing \ensuremath{\dot{M}} in this case.  Thus, examination of Figure~\ref{fig:eqmdot}d, for which $\gamma = -2/m$, indicates that the centrifugal force is lower for increasing \ensuremath{\dot{M}} (note that the hot plasma pressure current does not differ significantly between the different \ensuremath{\dot{M}} cases).  In Figure~\ref{fig:eqmod}d, on the other hand, in the inner region where the field strength decreases quickly, centrifugal force is larger for higher values of \ensuremath{\dot{M}}, while in the outer region, where the field is very weakly dependent on $\rho_e$, the centrifugal force is somewhat lower for higher values of \ensuremath{\dot{M}}.  Physically, the competing effects of increasing \ensuremath{\dot{M}}, i.e. increased plasma density but decreased angular velocity, mutually counteract in the middle magnetosphere, such that the magnetic field in this region becomes relatively insensitive to the value of \ensuremath{\dot{M}}.  It is important to note that the power law approximation is not perfectly applicable to the model results obtained here; for example, in the outer region, the field strength increases slowly with radial distance, a situation not considered by \cite{nichols03}, and for which the power law approximations were not designed.  Second in the outer region, the field does not map to a narrow band in the ionosphere, such that the the approximation conditions do not strictly hold in this region.  Hence, while caution should be used when comparing with the power law approximation, it nevertheless gives a reasonable insight into the behaviour of the system.  The profiles shown in Figures~\ref{fig:eqmdot}d and \ref{fig:eqmod}d also indicate why the ionospheric mapping differs between the two cases.  In the former case, the centrifugal force is solely dependent on the plasma angular velocity, which inside 15~\ensuremath{\mathrm{R_J}} is not particularly sensitive to \ensuremath{\dot{M}}, such that in this region the azimuthal current profiles, and thus the field mapping, do not differ greatly.  In the latter case, the centrifugal force also depends on the plasma density, such that the azimuthal current, and thus the field mapping, in the inner region varies significantly with \ensuremath{\dot{M}}.\\

Figures~\ref{fig:eqmdot}e and \ref{fig:eqmod}e indicate that taking $N_c \propto \ensuremath{\dot{M}}$ causes the cold plasma pressure to vary more significantly over the region inward of \ensuremath{\sim}60~\ensuremath{\mathrm{R_J}} than otherwise.  In addition, the \ensuremath{\dot{M}}~=~2000~\ensuremath{\mathrm{kg\;s^{-1}}} profile fits the \cite{frank02a} power laws best here, although it should be noted that these profiles are dependent on what reference value for \ensuremath{\dot{M}} is used and, for example, higher pressure values would be obtained if reference values less than 1000~\ensuremath{\mathrm{kg\;s^{-1}}} had been taken.\\

The azimuth-integrated radial current profiles are shown in Figures~\ref{fig:eqmdot}f and \ref{fig:eqmod}f, while the field-aligned current profiles are shown in Figures~\ref{fig:eqmdot}g and \ref{fig:eqmod}g.   Both sets of current profiles in Figures~\ref{fig:eqmdot} and \ref{fig:eqmod} are reasonably similar, differing most significantly in the degree to which they track the CAN-KK results, leading to different peak current values as will be discussed further below.  As \cite{nichols03} pointed out, for the CAN-KK field model the radial and field-aligned currents both tend to values dependent only on \ensuremath{\dot{M}} in the inner region, and the radial current tends to a value dependent on \ensuremath{\Sigma_P^*} at large distances.  In these figures the currents thus exhibit 3 distinct profiles in the inner region, in contrast to Figures~\ref{fig:eqsig}f and \ref{fig:eqsig}g, although the radial current profiles do not converge on a single value in the outer region due to the decrease in current intensity owing to the increased field strength over the CAN-KK model in this region. In both cases the field-aligned current reverses from upward to downward between \ensuremath{\sim}45-50~\ensuremath{\mathrm{R_J}}, decreasing with radial distance for higher values of \ensuremath{\dot{M}}.  \\

\begin{figure}
 \noindent\includegraphics[width=19pc]{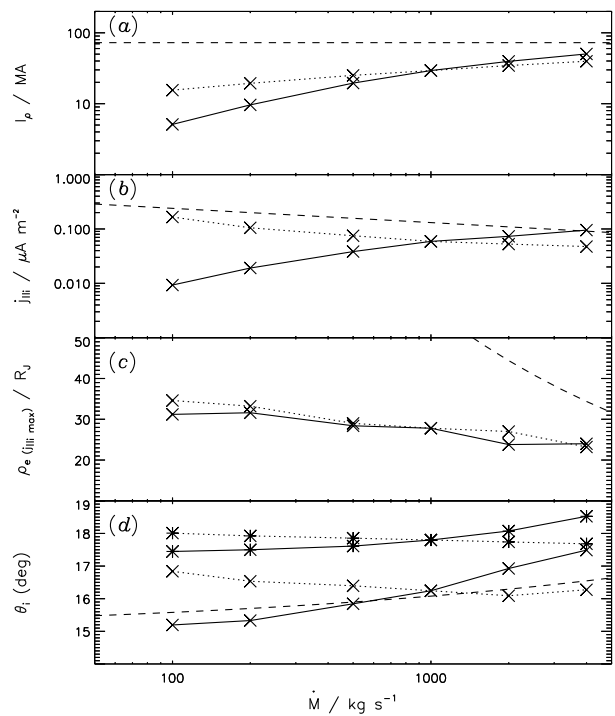}
\caption{
As for Figure~\ref{fig:maxsig}, except that here the results are plotted versus \ensuremath{\dot{M}}, and the points joined by the dotted lines assume the cold plasma density is independent of \ensuremath{\dot{M}}, while those joined by the solid lines assume it is given by equation~\ref{eq:ncprime}.  
}
\label{fig:maxmdot}
\end{figure}

Turning now to Figure~\ref{fig:maxmdot} we show the magnitudes and locations of the peak currents versus \ensuremath{\dot{M}} in the same format as for Figure~\ref{fig:maxsig}, except that here the points joined by the dotted lines show results taking the cold plasma density to be independent of \ensuremath{\dot{M}}, while those joined by the solid lines show those taking it to be given by equation~\ref{eq:ncprime}.  Figure~\ref{fig:maxmdot}a shows that the peak azimuth-integrated radial current increases with  \ensuremath{\dot{M}} for both cases, from \ensuremath{\sim}15~MA to \ensuremath{\sim}40~MA, and \ensuremath{\sim}5~MA to \ensuremath{\sim}50~MA for constant $N_c$ and $N_c \propto \ensuremath{\dot{M}}$, respectively as \ensuremath{\dot{M}} goes from 100~\ensuremath{\mathrm{kg\;s^{-1}}} to 4000~\ensuremath{\mathrm{kg\;s^{-1}}}, i.e.\ somewhat quicker for $N_c \propto \ensuremath{\dot{M}}$.  As is evident from Figure~\ref{fig:eqmdot}, this arises since, as \ensuremath{\dot{M}} increases while $N_c$ is constant, the magnetic field becomes less stretched due to the lower plasma angular velocity, such that the $I_\rho$ profiles fall away from the CAN-KK results at closer distances.  Thus, the peak currents increase slowly with \ensuremath{\dot{M}}.  On the other hand, for $N_c \propto \ensuremath{\dot{M}}$ the reverse is true, i.e.\ the field is more stretched for higher \ensuremath{\dot{M}}, such that the current profiles follow the CAN-KK results further, increasing the rate at which the peak current increases with \ensuremath{\dot{M}}.  This behaviour also accounts for the difference in the field-aligned current profiles shown in Figure~\ref{fig:maxmdot}b, in which \ensuremath{j_{\|i}} changes from \ensuremath{\sim}0.2~\ensuremath{\mu \mathrm{A\;m^{-2}}} to ~0.05~\ensuremath{\mu \mathrm{A\;m^{-2}}}, and \ensuremath{\sim}0.01~\ensuremath{\mu \mathrm{A\;m^{-2}}} to ~0.1~\ensuremath{\mu \mathrm{A\;m^{-2}}} for constant $N_c$ and $N_c \propto \ensuremath{\dot{M}}$, respectively as \ensuremath{\dot{M}} goes from 100~\ensuremath{\mathrm{kg\;s^{-1}}} to 4000~\ensuremath{\mathrm{kg\;s^{-1}}}.  \\

The equatorial distance of the peak field-aligned current shown in Figure~\ref{fig:maxmdot}c changes similarly for both cases, i.e.\ decreasing from \ensuremath{\sim}35~\ensuremath{\mathrm{R_J}} to \ensuremath{\sim}23~\ensuremath{\mathrm{R_J}}, and from \ensuremath{\sim}31~\ensuremath{\mathrm{R_J}} to \ensuremath{\sim}24~\ensuremath{\mathrm{R_J}} for constant $N_c$ and $N_c \propto \ensuremath{\dot{M}}$, respectively as \ensuremath{\dot{M}} goes from 100~\ensuremath{\mathrm{kg\;s^{-1}}} to 4000~\ensuremath{\mathrm{kg\;s^{-1}}}.  However, the difference in the field mapping results in the ionospheric co-latitudes plotted in Figure~\ref{fig:maxmdot}d varying differently with \ensuremath{\dot{M}} for the two cases.  First, as for Figure~\ref{fig:maxsig}d, for constant $N_c$ the radial motion of the peak field-aligned current is offset by the changing field mapping such that the ionospheric co-latitude of the peak current is only weakly dependent on \ensuremath{\dot{M}}, changing from from \ensuremath{\sim}16.8\ensuremath{^\circ{}} for \ensuremath{\dot{M}}~=~100~\ensuremath{\mathrm{kg\;s^{-1}}} to \ensuremath{\sim}16.3\ensuremath{^\circ{}} for \ensuremath{\dot{M}}~=~4000~\ensuremath{\mathrm{kg\;s^{-1}}}.  On the other hand, the reverse behaviour of the field mapping for $N_c \propto \ensuremath{\dot{M}}$ reinforces the radial motion in this case, such that the the co-latitude increases more rapidly than for the CAN-KK field, moving from \ensuremath{\sim}15.1\ensuremath{^\circ{}} for \ensuremath{\dot{M}}~=~100~\ensuremath{\mathrm{kg\;s^{-1}}} to \ensuremath{\sim}17.5\ensuremath{^\circ{}} for \ensuremath{\dot{M}}~=~4000~\ensuremath{\mathrm{kg\;s^{-1}}}.  Similarly the co-latitudes of the Ganymede footprint change from \ensuremath{\sim}18.0\ensuremath{^\circ{}} to \ensuremath{\sim}17.7\ensuremath{^\circ{}}, and \ensuremath{\sim}17.4\ensuremath{^\circ{}} to \ensuremath{\sim}18.5\ensuremath{^\circ{}} for constant $N_c$ and $N_c \propto \ensuremath{\dot{M}}$, respectively as \ensuremath{\dot{M}} goes from 100~\ensuremath{\mathrm{kg\;s^{-1}}} to 4000~\ensuremath{\mathrm{kg\;s^{-1}}}.  Thus, while it is difficult to generate the \ensuremath{\sim}3\ensuremath{^\circ{}} and \ensuremath{\sim}2\ensuremath{^\circ{}} shifts in latitude of the main oval and Ganymede footprint reported by \cite{grodent08}, these results suggest that a significant change in the iogenic plasma mass outflow rate, combined with an associated variation in the cold plasma density in the magnetosphere, possibly as a result of changing volcanic activity on Io, is the best candidate for explaining the shift in these auroral features.  In this case then, the blue image in Fig.~\ref{fig:g08} corresponds to an epoch of low volcanic activity, and the red image corresponds to an interval of high activity.


\section{Summary and Discussion}

In summary, we have considered the effect of a self-consistently computed magnetosdisc field structure on the magnetosphere-ionosphere coupling current system at Jupiter.  Specifically, we have incorporated the calculation of the plasma angular velocity profile using Hill-Pontius theory into the model of \cite{caudal86}, such that the resulting magnetosphere-ionosphere currents are computed using values of the equatorial magnetic field self-consistent with the plasma angular velocity profile.  We have thus obtained results using a more realistic plasma angular velocity profile than that used by \cite{caudal86}, and we have also included updated plasma parameters from Galileo data.  We have then examined the effect on the system of the values of two key magnetosphere-ionosphere coupling current system parameters, i.e.\ the ionospheric Pedersen conductivity \ensuremath{\Sigma_P^*} and iogenic plasma mass outflow rate \ensuremath{\dot{M}}.\\

We have thus found that the azimuthal current intensity is dependent on the values of \ensuremath{\Sigma_P^*} and \ensuremath{\dot{M}}.  Specifically, if the cold plasma density is taken to be independent of \ensuremath{\dot{M}}, we find that higher values of the quotient $(\ensuremath{\Sigma_P^*}/\ensuremath{\dot{M}})$ result in a more stretched magnetic field structure with a thinner, more intense current sheet due to the increased centrifugal force owing to higher plasma angular velocities.  We thus find that the results for $(\Sigma_P^* /\dot{M})$~=~$\ensuremath{\sim}10^{-4}$ to $5 \times 10^{-4}\; \mathrm{mho\;s\;kg^{-1}}$ provide the best agreement with various sets of observations, with both \ensuremath{|B_{ze}|} and $B_\varphi$ data most consistent with $(\Sigma_P^* /\dot{M})$~=~$5 \times 10^{-4}\; \mathrm{mho\;s\;kg^{-1}}$, and plasma angular velocity and pressure data in best agreement with $(\Sigma_P^* /\dot{M})$~=~$10^{-4}\; \mathrm{mho\;s\;kg^{-1}}$.  We discuss a possible reason for this discrepancy below.  In addition, we find that, while the equatorial azimuthal current in the inner region is dominated by hot plasma pressure, as is generally held to be the case at Jupiter, the use of a realistic plasma angular velocity profile actually results in the centrifugal current becoming dominant in the region beyond \ensuremath{\sim}35-50~\ensuremath{\mathrm{R_J}}, with the exact distance depending on the value of  $(\ensuremath{\Sigma_P^*}/\ensuremath{\dot{M}})$ taken.  This situation similar to that which has been determined for Saturn \citep{achilleos10a}. \\

Overall, the equatorial magnetic field profiles obtained are reasonably similar in the inner region to the  empirical CAN-KK model used in previous studies, such that the currents are of the same order as previous solutions obtained using this fixed equatorial field strength model. However, we show that the outer fringing field of the current disc acts to reverse the field-aligned current in the outer region, thus reproducing the dark region just poleward of the main oval.  The confinement of the upward current region to within \ensuremath{\sim}40-60~\ensuremath{\mathrm{R_J}} is consistent with the recent mapping of Jupiter's auroral features to the equatorial plane by \cite{vogt11a}.  These authors also determined the location of the open-closed field line boundary to be at \ensuremath{\sim}11\ensuremath{^\circ{}} co-latitude, a result which is also consistent with the \ensuremath{\sim}8-11\ensuremath{^\circ{}} co-latitudes of the last closed field line in the model presented here.  Further, we have found that, while the peak magnitudes of the M-I coupling currents are similar to those which have been determined previously, we have shown that the location of the peak currents differs significantly.  For example, the equatorial radial distance of the peak field-aligned current density increases with $(\Sigma_P^* /\dot{M})$ slower than simply using the CAN-KK model.  However, if the plasma density is independent of \ensuremath{\dot{M}}, this outward motion is counteracted by the simultaneous stretching of the field, such that the ionospheric co-latitude of the peak remains essentially constant.  We have therefore also examined the case whereby the plasma density is taken to be proportional to \ensuremath{\dot{M}}.  Hence, we found that in the inner region, where the field magnitude decreases quickly with distance, the centrifugal force increases with \ensuremath{\dot{M}}, while in the outer region, where the field in this model does not vary greatly with distance, the opposite is true.  Overall, the competing effects of increasing \ensuremath{\dot{M}}, i.e. increased plasma density but decreased angular velocity, mutually counteract in the middle magnetosphere, such that the magnetic field in this region is then relatively insensitive to the value of \ensuremath{\dot{M}}.  However, the nature of the centrifugal force in this case is such that changes to the field mapping induced by varying \ensuremath{\dot{M}} now reinforce the associated radial motion of the peak field-aligned current, such that the ionospheric co-latitude of the peak current varies more significantly, with higher values of \ensuremath{\dot{M}} corresponding to lower co-latitudes.  However, very large variations in the plasma mass outflow rate, well over an order of magnitude, are still required to reproduce shifts comparable to those observed by \cite{grodent08}.  \\

There are various directions in which this work should be taken forward.  First, the ionospheric Pedersen conductivity is assumed for simplicity to be constant, such that feedback effects due to auroral precipitation are neglected.  However, \cite{nichols04} showed that precipitation-induced enhancements of the Pedersen conductivity significantly affect the currents and plasma flows, such that this should be taken into account in future developments of this model.  Similarly, the field-aligned voltages required to drive the field-aligned currents are neglected in the present model.  The significance of such voltages has been previously debated \citep{nichols05,ray10a}, and it would be worth determining their effect in the model presented here.  Third, the \cite{caudal86} model neglects the effects of plasma pressure anisotropy, which has been shown by \cite{paranicas91a} to be a significant factor in the radial stress balance in Jupiter's  magnetosphere, and indeed has been recently shown to be important at Saturn \citep{kellett11a}.  It is probable that the omission of pressure anisotropy in the model is the cause of the discrepancy between the plasma and magnetic field data in terms of which value of $(\Sigma_P^* /\dot{M})$, i.e.\ $\ensuremath{\sim}10^{-4}$ or $5 \times 10^{-4}\; \mathrm{mho\;s\;kg^{-1}}$ provides the best agreement.  Development of the model to include this effect may produce significant inroads into the problem.  A fourth obvious area for further study is to examine the effect of the assumed sub-solar magnetopause distance.  Here we have simply taken the representative value of 85~\ensuremath{\mathrm{R_J}}, but observed values range over \ensuremath{\sim}45-100~\ensuremath{\mathrm{R_J}} \citep{khurana04a}, and \cite{khurana01} presented evidence of the solar wind's influence on Jupiter's magnetic field. \cite{caudal86} examined the effect of assumed magnetopause distance and showed that the larger the assumed distance, the more disc-like the field, and thus for different assumed magnetopause distances, the field mapping and M-I currents will be modified from those presented here, and this should be examined in future studies.  The location of the magnetopause is governed by the condition of pressure balance between a combination of magnetic and plasma pressures on one side and shocked solar wind ram pressure on the other, with variations typically being taken to be caused by variations in the solar wind dynamic pressure (e.g. \cite{huddleston98}).   \cite{cowley07} examined the effect of solar wind-induced expansions and compressions on the jovian M-I coupling current system using a prescribed field model.  In their model a strong compression which reduces \ensuremath{R_{mp}} from 85 to 45~\ensuremath{\mathrm{R_J}} results in modified field mapping such that the peak upward field-aligned current moves poleward by \ensuremath{\sim}1\ensuremath{^\circ{}}, and the ionospheric mapping of Ganymede's footprint at 15~\ensuremath{\mathrm{R_J}} shifts by \ensuremath{\sim}0.3\ensuremath{^\circ{}}  Thus, while the solar wind is expected to exert some influence on Jupiter's M-I coupling current system, these small latitudinal variations are not large enough to account for \emph{Grodent et al.'s}~[2008] observations.  It seems likely, therefore, that internal factors such as the iogenic plasma disc density are key, but it would be illuminating to determine the effect of the solar wind using the self-consistent model presented here.  Finally, although the present model is axisymmetric, the jovian current sheet is certainly not \citep{khurana04a}, such that the effect of this should be carefully examined in future studies.


\begin{acknowledgments}
	
	JDN was supported by STFC Grant ST/H002480/1, and acknowledges the support of ISSI, as this paper was discussed by ISSI International Team 178.  JDN also wishes to thank S.~W.~H.~Cowley for constructive comments during this study.  
	
\end{acknowledgments}
  
%
%

%
%

\end{article}



\begin{thebibliography}{53}
\providecommand{\natexlab}[1]{#1}
\expandafter\ifx\csname urlstyle\endcsname\relax
  \providecommand{\doi}[1]{doi:\discretionary{}{}{}#1}\else
  \providecommand{\doi}{doi:\discretionary{}{}{}\begingroup
  \urlstyle{rm}\Url}\fi

\bibitem[{\textit{Abramowitz and Stegun}(1965)}]{abramowitz65a}
Abramowitz, M., and I.~A. Stegun (1965), \textit{{Handbook of mathematical
  functions}}, Dover Publications Inc., Mineola, NY 11501, USA.

\bibitem[{\textit{Achilleos et~al.}(2010)\textit{Achilleos, Guio, and
  Arridge}}]{achilleos10a}
Achilleos, N., P.~Guio, and C.~S. Arridge (2010), {A model of force balance in
  Saturn's magnetodisc}, \textit{Mon. Not. R. Astron. Soc.}, \textit{401}(4),
  2349--2371, \doi{10.1111/j.1365-2966.2009.15865.x}.

\bibitem[{\textit{{Acu\~na} et~al.}(1983)\textit{{Acu\~na}, {Behannon}, and
  {Connerney}}}]{acuna83a}
{Acu\~na}, M.~H., K.~W. {Behannon}, and J.~E.~P. {Connerney} (1983), {Jupiter's
  magnetic field and magnetosphere}, in \textit{Physics of the Jovian
  Magnetosphere}, edited by {A.~J.~Dessler}, pp. 1--50, Cambridge. Univ. Press,
  Cambridge, UK.

\bibitem[{\textit{{Badman} et~al.}(2006)\textit{{Badman}, {Cowley}, {G\'erard},
  and {Grodent}}}]{badman06a}
{Badman}, S.~V., S.~W.~H. {Cowley}, J.-C. {G\'erard}, and D.~{Grodent} (2006),
  A statistical analysis of the location and width of saturn's southern
  auroras, \textit{Ann. Geophysicae}, \textit{24}(12), 3533--3545.

\bibitem[{\textit{Bagenal and Sullivan}(1981)}]{bagenal81a}
Bagenal, F., and J.~Sullivan (1981), {Direct plasma measurements in the Io
  torus and inner magnetosphere of Jupiter}, \textit{{J. Geophys. Res.}},
  \textit{86}, 8447--8466.

\bibitem[{\textit{{Bunce} et~al.}(2008)\textit{{Bunce}, {Arridge}, {Cowley},
  and {Dougherty}}}]{bunce08a}
{Bunce}, E.~J., C.~S. {Arridge}, S.~W.~H. {Cowley}, and M.~K. {Dougherty}
  (2008), {Magnetic field structure of Saturn's dayside magnetosphere and its
  mapping to the ionosphere: Results from ring current modeling}, \textit{{J.
  Geophys. Res.}}, \textit{113}, A02207, \doi{10.1029/2007JA012538}.

\bibitem[{\textit{{Caudal}}(1986)}]{caudal86}
{Caudal}, G. (1986), {A self-consistent model of {J}upiter's magnetodisc
  including the effects of centrifugal force and pressure}, \textit{J. Geophys.
  Res.}, \textit{91}, 4201--4221.

\bibitem[{\textit{Clarke et~al.}(2004)\textit{Clarke, {Grodent}, {Cowley},
  {Bunce}, {Zarka}, {Connerney}, and {Satoh}}}]{clarke04}
Clarke, J.~T., D.~{Grodent}, S.~W.~H. {Cowley}, E.~J. {Bunce}, P.~{Zarka},
  J.~E.~P. {Connerney}, and T.~{Satoh} (2004), {Jupiter's aurora}, in
  \textit{Jupiter.~The Planet, Satellites and Magnetosphere}, edited by
  {F.~Bagenal, T.~E.~Dowling and W.~B.~McKinnon}, pp. 639--670, Cambridge.
  Univ. Press, Cambridge, UK.

\bibitem[{\textit{{Connerney} et~al.}(1981)\textit{{Connerney}, {Acu\~na}, and
  {Ness}}}]{connerney81}
{Connerney}, J.~E.~P., M.~H. {Acu\~na}, and N.~F. {Ness} (1981), {Modeling the
  Jovian current sheet and inner magnetosphere}, \textit{J. Geophys. Res.},
  \textit{86}, 8370--8384.

\bibitem[{\textit{{Connerney} et~al.}(1998)\textit{{Connerney}, {Acu\~na},
  {Ness}, and {Satoh}}}]{connerney98}
{Connerney}, J.~E.~P., M.~H. {Acu\~na}, N.~F. {Ness}, and T.~{Satoh} (1998),
  {New models of Jupiter's magnetic field constrained by the Io flux tube
  footprint}, \textit{J. Geophys. Res.}, \textit{103}, 11,929--11,940,
  \doi{10.1029/97JA03726}.

\bibitem[{\textit{{Cowley} and {Bunce}}(2001)}]{cowley01}
{Cowley}, S.~W.~H., and E.~J. {Bunce} (2001), {Origin of the main auroral oval
  in Jupiter's coupled magnetosphere-ionosphere system}, \textit{Planet. Space
  Sci.}, \textit{49}, 1067--1088.

\bibitem[{\textit{{Cowley} et~al.}(2002)\textit{{Cowley}, {Nichols}, and
  {Bunce}}}]{cowley02}
{Cowley}, S.~W.~H., J.~D. {Nichols}, and E.~J. {Bunce} (2002), {Distributions
  of current and auroral precipitation in Jupiter's middle magnetosphere
  computed from steady-state Hill-Pontius angular velocity profiles: solutions
  for current sheet and dipole magnetic field models}, \textit{Planet. Space
  Sci.}, \textit{50}, 717--734.

\bibitem[{\textit{{Cowley} et~al.}(2003)\textit{{Cowley}, {Bunce}, and
  {Nichols}}}]{cowley03a}
{Cowley}, S.~W.~H., E.~J. {Bunce}, and J.~D. {Nichols} (2003), {Origins of
  Jupiter's main oval auroral emissions}, \textit{J. Geophys. Res.},
  \textit{108}, 8002, \doi{10.1029/2002JA009329}.

\bibitem[{\textit{{Cowley} et~al.}(2005)\textit{{Cowley}, {Alexeev},
  {Belenkaya}, {Bunce}, {Cottis}, {Kalegaev}, {Nichols}, {Prang\'e}, and
  {Wilson}}}]{cowley05a}
{Cowley}, S.~W.~H., I.~I. {Alexeev}, E.~S. {Belenkaya}, E.~J. {Bunce}, C.~E.
  {Cottis}, V.~V. {Kalegaev}, J.~D. {Nichols}, R.~{Prang\'e}, and F.~J.
  {Wilson} (2005), {A simple axisymmetric model of magnetosphere-ionosphere
  coupling currents in Jupiter's polar ionosphere}, \textit{J. Geophys. Res.},
  \textit{110}(A9), A11,209, \doi{10.1029/2005JA011237}.

\bibitem[{\textit{{Cowley} et~al.}(2007)\textit{{Cowley}, {Nichols}, and
  {Andrews}}}]{cowley07}
{Cowley}, S.~W.~H., J.~D. {Nichols}, and D.~J. {Andrews} (2007), {Modulation of
  Jupiter's plasma flow, polar currents, and auroral precipitation by solar
  wind-induced compressions and expansions of the magnetosphere: a simple
  theoretical model}, \textit{Ann. Geophysicae}, \textit{25}, 1433--1463.

\bibitem[{\textit{{Delamere} and {Bagenal}}(2003)}]{delamere03a}
{Delamere}, P.~A., and F.~{Bagenal} (2003), {Modeling variability of plasma
  conditions in the Io torus}, \textit{J. Geophys. Res.}, \textit{108(A7)},
  1276, \doi{10.1029/2002JA009706}.

\bibitem[{\textit{{Dols} et~al.}(2008)\textit{{Dols}, {Delamere}, and
  {Bagenal}}}]{dols08a}
{Dols}, V., P.~A. {Delamere}, and F.~{Bagenal} (2008), {A multispecies
  chemistry model of Io's local interaction with the Plasma Torus},
  \textit{J.~Geophys.~Res.}, \textit{113}(A9), A09208,
  \doi{10.1029/2007JA012805}.

\bibitem[{\textit{{Frank} et~al.}(2002)\textit{{Frank}, {Paterson}, and
  {Khurana}}}]{frank02a}
{Frank}, L.~A., W.~R. {Paterson}, and K.~K. {Khurana} (2002), {Observations of
  thermal plasmas in Jupiter's magnetotail}, \textit{{J. Geophys. Res.}},
  \textit{107}(A1), 1003, \doi{10.1029/2001JA000077}.

\bibitem[{\textit{{Grodent} et~al.}(2003{\natexlab{a}})\textit{{Grodent},
  Clarke, {Kim}, {Waite}, and {Cowley}}}]{grodent03b}
{Grodent}, D., J.~T. Clarke, J.~{Kim}, J.~H. {Waite}, and S.~W.~H. {Cowley}
  (2003{\natexlab{a}}), {Jupiter's main auroral oval observed with HST-STIS},
  \textit{J. Geophys. Res.}, \textit{108}, 1389, \doi{10.1029/2003JA009921}.

\bibitem[{\textit{{Grodent} et~al.}(2003{\natexlab{b}})\textit{{Grodent},
  Clarke, {Waite}, {Cowley}, {G\'erard}, and {Kim}}}]{grodent03a}
{Grodent}, D., J.~T. Clarke, J.~H. {Waite}, S.~W.~H. {Cowley}, J.-C.
  {G\'erard}, and J.~{Kim} (2003{\natexlab{b}}), {Jupiter's polar auroral
  emissions}, \textit{J. Geophys. Res.}, \textit{108}, 1366,
  \doi{10.1029/2003JA010017}.

\bibitem[{\textit{{Grodent} et~al.}(2008)\textit{{Grodent}, {G\'erard},
  {Radioti}, {Bonfond}, and {Saglam}}}]{grodent08}
{Grodent}, D., J.-C. {G\'erard}, A.~{Radioti}, B.~{Bonfond}, and A.~{Saglam}
  (2008), {Jupiter's changing auroral location}, \textit{J. Geophys. Res.},
  \textit{113}(A12), A01,206, \doi{10.1029/2007JA012601}.

\bibitem[{\textit{{Hill}}(1979)}]{hill79}
{Hill}, T.~W. (1979), {Inertial limit on corotation}, \textit{J. Geophys.
  Res.}, \textit{84}, 6554--6558.

\bibitem[{\textit{{Hill}}(1980)}]{hill80}
{Hill}, T.~W. (1980), {Corotation lag in Jupiter's magnetosphere - Comparison
  of observation and theory}, \textit{Science}, \textit{207}, 301.

\bibitem[{\textit{{Hill}}(2001)}]{hill01}
{Hill}, T.~W. (2001), {The Jovian auroral oval}, \textit{J. Geophys. Res.},
  \textit{106}, 8101--8108, \doi{10.1029/2000JA000302}.

\bibitem[{\textit{{Huang} and {Hill}}(1989)}]{huang89}
{Huang}, T.~S., and T.~W. {Hill} (1989), {Corotation lag of the Jovian
  atmosphere, ionosphere, and magnetosphere}, \textit{J. Geophys. Res.},
  \textit{94}, 3761--3765.

\bibitem[{\textit{{Huddleston} et~al.}(1998)\textit{{Huddleston}, {Russell},
  {Kivelson}, {Khurana}, and {Bennett}}}]{huddleston98}
{Huddleston}, D.~E., C.~T. {Russell}, M.~G. {Kivelson}, K.~K. {Khurana}, and
  L.~{Bennett} (1998), {Location and shape of the Jovian magnetopause and bow
  shock}, \textit{J. Geophys. Res.}, \textit{103}, 20,075--20,082,
  \doi{10.1029/98JE00394}.

\bibitem[{\textit{{Kane} et~al.}(1995)\textit{{Kane}, {Mauk}, {Keath}, and
  {Krimigis}}}]{kane95}
{Kane}, M., B.~H. {Mauk}, E.~P. {Keath}, and S.~M. {Krimigis} (1995), {Hot ions
  in Jupiter's magnetodisc: A model for Voyager 2 low-energy charged particle
  measurements}, \textit{J. Geophys. Res.}, \textit{100}, 19,473--19,486,
  \doi{10.1029/95JA00793}.

\bibitem[{\textit{Kellett et~al.}(2011)\textit{Kellett, Arridge, Bunce, Coates,
  Cowley, Dougherty, Persoon, Sergis, and Wilson}}]{kellett11a}
Kellett, S., C.~S. Arridge, E.~J. Bunce, A.~J. Coates, S.~W.~H. Cowley, M.~K.
  Dougherty, A.~M. Persoon, N.~Sergis, and R.~J. Wilson (2011), {Saturn's ring
  current: Local time dependence and temporal variability}, \textit{J. Geophys.
  Res.}, \textit{116}, A05220, \doi{10.1029/2010JA016216}.

\bibitem[{\textit{{Khurana}}(2001)}]{khurana01}
{Khurana}, K.~K. (2001), {Influence of solar wind on Jupiter's magnetosphere
  deduced from currents in the equatorial plane}, \textit{J. Geophys. Res.},
  \textit{106}, 25,999--26,016, \doi{10.1029/2000JA000352}.

\bibitem[{\textit{{Khurana} and {Kivelson}}(1993)}]{khurana93}
{Khurana}, K.~K., and M.~G. {Kivelson} (1993), {Inference of the angular
  velocity of plasma in the Jovian magnetosphere from the sweepback of magnetic
  field}, \textit{J. Geophys. Res.}, \textit{98(A1)}, 67--79.

\bibitem[{\textit{{Khurana} et~al.}(2004)\textit{{Khurana}, {Kivelson},
  {Vasyli{\=u}nas}, {Krupp}, {Woch}, {Lagg}, {Mauk}, and {Kurth}}}]{khurana04a}
{Khurana}, K.~K., M.~G. {Kivelson}, V.~M. {Vasyli{\=u}nas}, N.~{Krupp},
  J.~{Woch}, A.~{Lagg}, B.~H. {Mauk}, and W.~S. {Kurth} (2004), {The
  configuration of Jupiter's magnetosphere}, in \textit{Jupiter.~The Planet,
  Satellites and Magnetosphere}, edited by {F.~Bagenal, T.~E.~Dowling and
  W.~B.~McKinnon}, pp. 593--616, Cambridge. Univ. Press.

\bibitem[{\textit{{Krimigis} et~al.}(1981)\textit{{Krimigis}, {Carbary},
  {Keath}, {Bostrom}, {Axford}, {Gloeckler}, {Lanzerotti}, and
  {Armstrong}}}]{krimigis81}
{Krimigis}, S.~M., J.~F. {Carbary}, E.~P. {Keath}, C.~O. {Bostrom}, W.~I.
  {Axford}, G.~{Gloeckler}, L.~J. {Lanzerotti}, and T.~P. {Armstrong} (1981),
  {Characteristics of hot plasma in the Jovian magnetosphere - Results from the
  Voyager spacecraft}, \textit{J. Geophys. Res.}, \textit{86}, 8227--8257.

\bibitem[{\textit{Lackner}(1970)}]{lackner70a}
Lackner, K. (1970), {Deformation of a magnetic dipole field by trapped
  particles}, \textit{{J. Geophys. Res.}}, \textit{75}(16), 3180--3192.

\bibitem[{\textit{{Mauk} and {Krimigis}}(1987)}]{mauk87a}
{Mauk}, B.~H., and S.~M. {Krimigis} (1987), {Radial force balance within
  Jupiter's dayside magnetosphere}, \textit{{J. Geophys. Res.}},
  \textit{92}(A9), 9931--9941.

\bibitem[{\textit{{Mauk} and {Saur}}(2007)}]{mauk07a}
{Mauk}, B.~H., and J.~{Saur} (2007), {Equatorial electron beams and auroral
  structuring at Jupiter}, \textit{{J. Geophys. Res.}}, \textit{112}, A10221,
  \doi{10.1029/2007JA012370}.

\bibitem[{\textit{{McNutt}}(1983)}]{mcnutt83a}
{McNutt}, R.~L., Jr. (1983), {Force balance in the magnetospheres of Jupiter
  and Saturn}, \textit{Adv. Space Res.}, \textit{3}(3), 55--58.

\bibitem[{\textit{{McNutt}}(1984)}]{mcnutt84a}
{McNutt}, R.~L., Jr. (1984), {Force Balance in Outer Planet Magnetospheres}, in
  \textit{{Physics of Space Plasmas, Proceedings of the 1982-4 MIT Symposia.
  SPI Conference Proceedings and Reprint Series}}, vol.~5, edited by
  {J.~Belcher, H.~Bridge, T.~Change, B.~Coppi, \& J.~R.~Jasperse}, pp.
  179--210, Scientific Publishers, Cambridge, Mass.

\bibitem[{\textit{{McNutt} et~al.}(1981)\textit{{McNutt}, Belcher, and
  Bridge}}]{mcnutt81a}
{McNutt}, R.~L., Jr., J.~W. Belcher, and H.~S. Bridge (1981), {Positive-ion
  observations in the middle magnetosphere of Jupiter}, \textit{{J. Geophys.
  Res.}}, \textit{86}(NA10), 8319--8342.

\bibitem[{\textit{{Millward} et~al.}(2005)\textit{{Millward}, {Miller},
  {Stallard}, {Achilleos}, and {Aylward}}}]{millward05}
{Millward}, G., S.~{Miller}, T.~{Stallard}, N.~{Achilleos}, and A.~D. {Aylward}
  (2005), {On the dynamics of the jovian ionosphere and thermosphere IV:
  ion-neutral coupling}, \textit{Icarus}, \textit{173}, 200--211,
  \doi{10.1016/j.icarus.2004.07.027}.

\bibitem[{\textit{{Nichols}}(2011)}]{nichols11a}
{Nichols}, J.~D. (2011), {Magnetosphere-ionosphere coupling at Jupiter-like
  exoplanets with internal plasma sources: implications for detectability of
  auroral radio emissions}, \textit{Mon. Not. R. Astron. Soc.}, \textit{414},
  2125---2138, \doi{10.1111/j.1365-2966.2011.18528.x}.

\bibitem[{\textit{{Nichols} and {Cowley}}(2003)}]{nichols03}
{Nichols}, J.~D., and S.~W.~H. {Cowley} (2003), {Magnetosphere-ionosphere
  coupling currents in Jupiter's middle magnetosphere: dependence on the
  effective ionospheric Pedersen conductivity and iogenic plasma mass outflow
  rate}, \textit{Ann. Geophysicae}, \textit{21}, 1419--1441.

\bibitem[{\textit{{Nichols} and {Cowley}}(2004)}]{nichols04}
{Nichols}, J.~D., and S.~W.~H. {Cowley} (2004), {Magnetosphere-ionosphere
  coupling currents in Jupiter's middle magnetosphere: effect of
  precipitation-induced enhancement of the ionospheric Pedersen conductivity},
  \textit{Ann. Geophysicae}, \textit{22}, 1799--1827.

\bibitem[{\textit{{Nichols} and {Cowley}}(2005)}]{nichols05}
{Nichols}, J.~D., and S.~W.~H. {Cowley} (2005), {Magnetosphere-ionosphere
  coupling currents in Jupiter's middle magnetosphere: effect of
  magnetosphere-ionosphere decoupling by field-aligned auroral voltages},
  \textit{Ann. Geophysicae}, \textit{23}, 799--808.

\bibitem[{\textit{{Nichols} et~al.}(2009)\textit{{Nichols}, Clarke, {G\'erard},
  {Grodent}, and {Hansen}}}]{nichols09b}
{Nichols}, J.~D., J.~T. Clarke, J.-C. {G\'erard}, D.~{Grodent}, and K.~C.
  {Hansen} (2009), {Variation of different components of Jupiter's auroral
  emission}, \textit{{J. Geophys. Res.}}, \textit{114}, A06210,
  \doi{10.1029/2009JA014051}.

\bibitem[{\textit{Paranicas et~al.}(1991)\textit{Paranicas, Mauk, and
  Krimigis}}]{paranicas91a}
Paranicas, C.~P., B.~H. Mauk, and S.~M. Krimigis (1991), {Pressure Anisotropy
  and Radial Stress Balance in the Jovian Neutral Sheet}, \textit{J. Geophys.
  Res.}, \textit{96}(A12), 21,135--21,140.

\bibitem[{\textit{{Pontius}}(1997)}]{pontius97}
{Pontius}, D.~H. (1997), {Radial mass transport and rotational dynamics},
  \textit{J. Geophys. Res.}, \textit{102}, 7137--7150, \doi{10.1029/97JA00289}.

\bibitem[{\textit{{Ray} et~al.}(2010)\textit{{Ray}, {Ergun}, {Delamere}, and
  {Bagenal}}}]{ray10a}
{Ray}, L.~C., R.~E. {Ergun}, P.~A. {Delamere}, and F.~{Bagenal} (2010),
  {Magnetosphere-ionosphere coupling at Jupiter: Effect of field-aligned
  potentials on angular momentum transport}, \textit{{J. Geophys. Res.}},
  \textit{115}, A09211,, \doi{10.1029/2010JA015423}.

\bibitem[{\textit{{Siscoe} and {Summers}}(1981)}]{siscoe81}
{Siscoe}, G.~L., and D.~{Summers} (1981), {Centrifugally driven diffusion of
  Iogenic plasma}, \textit{J. Geophys. Res.}, \textit{86}, 8471--8479.

\bibitem[{\textit{{Smith} and {Aylward}}(2009)}]{smith09a}
{Smith}, C. G.~A., and A.~D. {Aylward} (2009), Coupled rotational dynamics of
  jupiter's thermosphere and magnetosphere, \textit{Ann. Geophysicae},
  \textit{27}(1), 199--230.

\bibitem[{\textit{Tao et~al.}(2009)\textit{Tao, Fujiwara, and Kasaba}}]{tao09a}
Tao, C., H.~Fujiwara, and Y.~Kasaba (2009), Neutral wind control of the jovian
  magnetosphere-ionosphere current system, \textit{J. Geophys. Res.},
  \textit{114}(A8), A08307, \doi{10.1029/2008JA013966}.

\bibitem[{\textit{{Tao} et~al.}(2010)\textit{{Tao}, {Fujiwara}, and
  {Kasaba}}}]{tao10a}
{Tao}, C., H.~{Fujiwara}, and Y.~{Kasaba} (2010), {Jovian
  magnetosphere--ionosphere current system characterized by diurnal variation
  of ionospheric conductance}, \textit{Planet. Space Sci.}, \textit{58},
  351--364, \doi{10.1016/j.pss.2009.10.005}.

\bibitem[{\textit{{Vasyli{\=u}nas}}(1983)}]{vasyliunas83}
{Vasyli{\=u}nas}, V.~M. (1983), {Plasma distribution and flow}, in
  \textit{Physics of the Jovian Magnetosphere}, edited by {A.~J.~Dessler}, pp.
  395--453, Cambridge. Univ. Press, Cambridge, UK.

\bibitem[{\textit{Vogt et~al.}(2011)\textit{Vogt, Kivelson, Khurana, Walker,
  Bonfond, Grodent, and Radioti}}]{vogt11a}
Vogt, M.~F., M.~G. Kivelson, K.~K. Khurana, R.~J. Walker, B.~Bonfond,
  D.~Grodent, and A.~Radioti (2011), {Improved mapping of Jupiter's auroral
  features to magnetospheric sources}, \textit{J. Geophys. Res.},
  \textit{116}(A3), A03220, \doi{10.1029/2010JA016148}.

\end{thebibliography}
\end{document}